\documentclass[aps,pra,twocolumn,floatfix,showpacs,lengthcheck,citeautoscript,superscriptaddress]{revtex4-1}
\usepackage{graphicx}
\usepackage{amsmath}
\usepackage{amsfonts}
\usepackage{mathrsfs}
\usepackage{color}

\usepackage{amssymb}
\usepackage{bm}
\usepackage{times}
\usepackage{MnSymbol}

\begin{document}

\title{Spectral fingerprints of the non-linear dynamics of driven superconductors with dissipation}

\author{H. P. {Ojeda Collado}} 
\affiliation{Centro At{\'{o}}mico Bariloche and Instituto Balseiro,
Comisi\'on Nacional de Energ\'{\i}a At\'omica (CNEA)--Universidad Nacional de Cuyo (UNCUYO), 8400 Bariloche, Argentina}
\affiliation{Instituto de Nanociencia y Nanotecnolog\'{i}a (INN), Consejo Nacional de Investigaciones Cient\'{\i}ficas y T\'ecnicas (CONICET)--CNEA, 8400 Bariloche, Argentina}
\author{Gonzalo Usaj}
\affiliation{Centro At{\'{o}}mico Bariloche and Instituto Balseiro,
Comisi\'on Nacional de Energ\'{\i}a At\'omica (CNEA)--Universidad Nacional de Cuyo (UNCUYO), 8400 Bariloche, Argentina}
\affiliation{Instituto de Nanociencia y Nanotecnolog\'{i}a (INN), Consejo Nacional de Investigaciones Cient\'{\i}ficas y T\'ecnicas (CONICET)--CNEA, 8400 Bariloche, Argentina}
\author{Jos\'{e} Lorenzana}
\affiliation{ISC-CNR and Department of Physics, Sapienza University of Rome, Piazzale Aldo Moro 2, I-00185, Rome, Italy}
\author{C. A. Balseiro}
\affiliation{Centro At{\'{o}}mico Bariloche and Instituto Balseiro,
Comisi\'on Nacional de Energ\'{\i}a At\'omica (CNEA)--Universidad Nacional de Cuyo (UNCUYO), 8400 Bariloche, Argentina}
\affiliation{Instituto de Nanociencia y Nanotecnolog\'{i}a (INN), Consejo Nacional de Investigaciones Cient\'{\i}ficas y T\'ecnicas (CONICET)--CNEA, 8400 Bariloche, Argentina}

\begin{abstract}
We study the out-of-equilibrium dynamics of a BCS superconductor in the presence of a periodic drive with a frequency $\omega_d$ larger than the equilibrium superconducting gap and an external bath providing dissipation. Similar to the dissipationless case, a subset of quasiparticles, resonant with the drive, synchronizes and produces interesting non-linear phenomena. For small dissipation Rabi-Higgs oscillations [Collado et al. Phys. Rev. B {\bf 98},  214519  (2018)] can be observed as a transient effect. At long-times in contrast, dissipation leads to a steady state with a population imbalance that increases quadratically for small drives and saturates for large drives as in non-linear quantum optics. We show also that second harmonic generation is allowed for a drive which acts on the BCS coupling constant. We also compute the intensity of time- and angle-resolved photo-emission spectroscopy (tr-ARPES) and time-resolved tunneling spectra. The excited quasiparticle population appears as a decrease (increase) in the photo-emission intensity at energy $-\omega_d/2$ ($+\omega_d/2$) measured from the chemical potential. The tunneling intensity shows a time-dependent depression at $\pm\omega_d/2$ due to the lacking of spectral density and population unbalance causing Pauli blockade. We predict that at short times, compared to the energy relaxation time, both experiments will show oscillations with the Rabi-Higgs frequency.  
\end{abstract}
\date{\today}
\maketitle

\section{Introduction}

The recent advances in laser technologies have opened new avenues for the study of collective behavior and emergent phenomena in condensed matter \cite{Fausti2011d,Mansart2013,Matsunaga2013,Matsunaga2014,Mankowsky2014,Nicoletti2014,Kaiser2014,Mitrano2016,Rajasekaran2018} and ultracold-atom systems \cite{Stoferle2004,Haller2010,Endres2012,Chin2010,Behrle2018,Clark2015} far from the linear response paradigm. Theoretical examples of these strongly non-linear phenomena are dynamical phase transitions after a quantum quench \cite{Barankov2006a,Eckstein2009} or under a periodic drive \cite{Collado2018}.

While these studies are of interest per se, they can also, in principle, shine some light on the nature of the precursor equilibrium phases. In this context, superconducting condensates have drawn increasing attention, both in theoretical and experimental research, as challenging cases for the study of collective out of equilibrium states, especially in the case of superconductors with competing orders \cite{Fausti2011d,Mansart2013,Matsunaga2013,Matsunaga2014,Mankowsky2014,Nicoletti2014,Kaiser2014,Mitrano2016,Rajasekaran2018}.
Much of the interest in the field was fueled by theoretical studies of quenched systems \cite{Volkov1974,Barankov2006a,Barankov2004}
and experiments in which the superconductor is excited by a pump pulse without a complete suppression of the superconducting state. In the perturbed system, the superconducting order parameter $\Delta$ (Higgs mode) or the charge modes evolve with collective oscillations at frequency $2\Delta$  which are rapidly damped due to dephasing \cite{Mansart2013,Matsunaga2013}.

Another interesting way to manipulate a condensate is through periodic drives. In solid-state superconductors, different type of drives have been discussed \cite{Collado2018}. Among them are impulsive stimulated Raman scattering (ISRS) \cite{Lorenzana2013}, phonon assisted modulation of the density of states (DOS-driving) as  in the case of 2H-NbSe$_2$ \cite{Balseiro1980,Littlewood1982} or of the coupling constant ($\lambda$-driving) as proposed for FeSe \cite{Collado2018}, microwave drives and THz drives \cite{Cea2015} as already realized in Ref.~[\onlinecite{Matsunaga2014}]. Periodic drives can also be achieved in ultra-cold atoms where there are well known techniques to modify Hamiltonian parameters at will: DOS-driving can be achieved modifying the depth of a periodic potential, as it has been done for bosons \cite{Stoferle2004,Haller2010,Endres2012}, while $\lambda$-driven can be implemented in a variety of ways \cite{Chin2010,Behrle2018,Clark2015}.

Previously, we have shown that within a BCS self-consistent dynamics, a periodic excitation at a frequency $\omega_d$ in resonance with quasiparticle excitations ($\omega_d>2\Delta$) can produce collective Rabi oscillations (“Rabi-Higgs” mode) of the quasiparticle population with a frequency proportional to the strength of the drive \cite{Collado2018}. This is due to a subset of quasiparticle excitations with energy $E_{\bm{k}}$ satisfying $\omega_d\approx 2 E_{\bm{k}}=2\sqrt{\xi_{\bm{k}}^2+\Delta^2}$ where $E_{\bm{k}}$ is the quasiparticle energy and $\xi_{\bm{k}}$ is the energy of the fermions measured from the chemical potential. A family of quasiparticles approximately satisfying the above condition synchronize among themselves and perform collective oscillations at the Rabi frequency.

The observation of a Rabi-Higgs mode requires entering into a highly nonlinear regime where damping and decoherence effects can be overcome. It is thus interesting to discuss under which conditions this non-linear regime can be achieved. It is well known from optical Bloch equations for a periodically driven two-level system \cite{Steck2019} that a finite energy relaxation time $\tau$ leads to a linear response regime at long times. The non-linear regime of  Rabi oscillations can be accessed at short times when the  Rabi frequency is larger than $1/\tau$. Similarly, the observation of Rabi-Higgs oscillations in a superconductor requires long relaxation times and/or strong drives (which result in Rabi-Higgs frequency $\Omega_R>1/\tau$). We mention that a similar phenomena has been proposed for driven graphene \cite{Mishchenko2009} where, however, the physics is much simpler as quasiparticle interactions have not been considered.

How long can $\tau$ be? Quasiparticle relaxation times in superconductors can be extremely long under favorable conditions. A simple estimate can be obtained through  the Dynes parameter \cite{Dynes1978} in tunneling experiments which, as explicitly shown in Ref.~[\onlinecite{Collado2019}], is directly related to the damping of quasiparticles by the bath ultimately limiting the coherent dynamics. In aluminium samples, an inverse Dynes parameter $1/\gamma \equiv \tau\sim 10^{6}/\Delta\sim \mu$s has been measured  \cite{Saira2012} which suggest that there is a large time window for coherent dynamics. Indeed, in Ref.~[\onlinecite{Collado2018}] we found that for a moderate drive strength the Rabi period $\tau_R\equiv 2\pi/\Omega_R$  is in the scale of tenths or hundreds of $1/\Delta\ll \tau$ (see also Fig.~\ref{fig:fig1} below). On the other hand, the above estimate for $\tau$ is probably too optimistic as the same out-of-equilibrium quasiparticles will open new relaxation channels \cite{Chang1978} and hence the whole out-of-equilibrium many-body problem can be considered.

Here we take a simple approach for the interplay of coherent non-linear phenomena and damping by considering a driven superconductor in the presence of a bath providing energy relaxation. We make the simplest possible assumption for the bath leaving a microscopic description for future work. Thus we treat the coupling with the bath as a phenomenological parameter and   
present the spectral signatures  of the Rabi-Higgs modes varying the coupling and other experimental parameters. Differently from previous works, the  dissipative dynamics is treated in a self-consistent way using the method introduced in Ref.~[\onlinecite{Collado2019}]. Both tr-ARPES and tunneling are analyzed in detail.

\section{The time-dependent BCS model with dissipation}

In the absence of dissipation, $\lambda$-drive and DOS-drive produce qualitatively similar results \cite{Collado2018}. As we expect the same to be true in the presence of dissipation, we restrict hereon to study $\lambda$-driving. 

In addition, in the dissipationless case, even for weak drives and for $\omega_d>2\Delta$ the Rabi-Higgs mode appears with  frequency $\Omega_R$ proportional to the intensity of the drive. Apparently the Rabi-Higgs mode violates any linear response prescription but, as mention in the introduction, without dissipation the system becomes inherently non-linearly at long times no matter how weak the perturbation is. In other words, the order of limits is important. Taking first the Rabi frequency  (proportional to the drive intensity) $\Omega_R\rightarrow 0$ and then energy relaxation times $\tau\rightarrow \infty$ a linear response regime is well defined. Inverting the order of limits it is not. In the following we study numerically the effect of dissipation on the Rabi-Higgs response. 

\subsection{Model and formalism}
We consider a single-band s-wave superconductor described by the Hamiltonian
\begin{equation}
\label{eq:HBCS}
H_{\mathrm{BCS}}=\sum_{\bm{k},\sigma}\xi_{\bm{k}}c_{\bm{k}\sigma}^{\dagger}c_{\bm{k}\sigma}^{}-\lambda(t)\sum_{\bm{k},\bm{k^{\prime}}}c_{\bm{k}\uparrow}^{\dagger}c_{\bm{-k}\downarrow}^{\dagger}c_{\bm{-k^{\prime}}\downarrow}^{}c_{\bm{k^{\prime}}\uparrow}^{}
\end{equation}
where $c_{\bm{k}\sigma}$ ($c_{\bm{k}\sigma}^{\dagger})$
destroys (creates) an electron with momentum $\bm{k}$, energy $\varepsilon_{\bm{k}}$ and spin
$\sigma$. Here  $\xi_{\bm{k}}=\varepsilon_{\bm{k}}-\mu$ measures the energy from the Fermi level $\mu$ and the pairing interaction $\lambda(t)$ is parameterized as
\begin{equation}
\label{eq:ldt}
\lambda(t)=\lambda_0[1+\Theta\left(t\right)\alpha\sin\left(\omega_d t\right)]\,,
\end{equation}
where $\Theta\left(t\right)$ is the Heaviside step function. In most of our calculations we take the parameter $\alpha\in\left[0,0.2\right]$, that corresponds to a modulation of up to $20$\% of the equilibrium pairing interaction $\lambda_0$, to keep it within the range of experimental accessibility \cite{Collado2018,Sentef2016}. In the thermodynamic limit, the Hamiltonian~(\ref{eq:HBCS}) is equivalent to the mean-field Hamiltonian,
\begin{eqnarray}
  \label{eq:hmf}
H_{\mathrm{MF}}=\sum_{\bm{k}}\psi_{\bm{k}}^{\dagger}\bm{H}_{\bm{k}}(t)\psi_{\bm{k}}, 
\end{eqnarray}
written in the Nambu spinor basis $\psi_{\bm{k}}=\left(c_{\bm{k}\uparrow},c_{-\bm{k}\downarrow}^{\dagger}\right)^{\mathrm{T}}$, where
\begin{equation}
\label{eq:hkt}
\bm{H}_{\bm{k}}(t)=\left(\begin{array}{cc}
\xi_{\bm{k}} & -\Delta(t)\\
-\Delta(t)^{*} & -\xi_{\bm{k}}
\end{array}\right)\,,
\end{equation}
and the instantaneous superconducting order parameter is given by
\begin{equation}
\label{eq:ssd}
\Delta(t)=\lambda(t)\sum_{\bm{k}}\left\langle c_{\bm{k}\uparrow}^{\dagger}(t)c_{\bm{-k}\downarrow}^{\dagger}(t)\right\rangle .
\end{equation}
Here $\left\langle \ldots\right\rangle $ denotes the expectation value on the initial state. 

In order to consider dissipation we couple the  superconductor to a reservoir.
For simplicity we will take the bath to be at zero temperature, which means that if the superconductor is out of equilibrium at some time and it is allowed to evolve in the absence of the drive, it will eventually relax to the ground state with the bath absorbing all the excess energy. The method to treat the bath was explained in detail in Ref.~[\onlinecite{Collado2019}]. Here we summarize the main results.
 
To describe the reservoir effect, the self-consistent solution of the gap equation is written in terms of the Keldysh two-time contour Green\textquoteright{}s functions. In the Nambu spinor basis, the retarded and lesser Green functions are $2\times 2$ matrices with matrix elements given by
\begin{eqnarray}
\label{eq:greendeff}
\notag
\bm{G}_{\bm{k}}^{R}\left(t,t^{\prime}\right)_{\alpha\beta}&=&-i\Theta\left(t-t^{\prime}\right)\left\langle \left\{ \psi_{\bm{k}{\alpha}}(t),\psi_{\bm{k}{\beta}}^{\dagger}\left(t^{\prime}\right)\right\} \right\rangle,\\
\bm{G}_{\bm{k}}^{<}\left(t,t^{\prime}\right)_{\alpha\beta}&=&i\left\langle \psi_{\bm{k}{\alpha}}^{\dagger}\left(t^{\prime}\right)\psi_{\bm{k}{\beta}}(t)\right\rangle, 
\end{eqnarray}
respectively. Thus, the superconducting gap equation [Eq.~(\ref{eq:ssd})] can be written as
\begin{equation} 
\label{eq:deltadef}
\Delta(t)=-i \lambda(t) \sum_{\bm{k}} {\bm{G}_{\bm{k}}^{<}\left(t,t\right)}_{12}\,. 
\end{equation}
When considering the coupling to a reservoir, the lesser Green function satisfies the Keldysh equation in time domain \cite{Antipekka1994,Horacio1992,Moore2019,Collado2019}
\begin{equation}
\label{eq:gl}
{\bm{G}}_{\bm{k}}^{<}(t,t^{\prime})=\int dt_{1}\int dt_{2}\,{\bm{G}}_{\bm{k}}^{R}(t,t_{1}){\bm{\Sigma}}_{\bm{k}}^{<}(t_{1},t_{2}){\bm{G}}_{\bm{k}}^{R}(t^{\prime},t_{2})^{\dagger}\,
\end{equation}
where the dissipation effects are taken into account via the lesser self-energy ${\bm{\Sigma}}_{\bm{k}}^{<}(t_{1},t_{2})$ and the retarded Green function which is solution of the corresponding Dyson equation with a retarded self-energy ${\bm{\Sigma}}_{\bm{k}}^{R}(t_{1},t_{2})$.
Following Refs.~\cite{Moore2019,Millis2017} we consider a mechanism for dissipation that couples each pair of states $\bm{k}\uparrow, -\bm{k}\downarrow$ with a reservoir described by a time-independent one body Hamiltonian  $H_{b}=\sum_{\ell,\sigma}E_{\ell}d_{\ell\sigma}^{\dagger}d_{\ell\sigma}$ where $d_{\ell\sigma}^{\dagger}$ creates a state in a single particle bath state with energy $E_{\ell}$. In the limit of a wide-band reservoir with identical coupling $V_{\bm{k}\ell}=V_\ell$ for each $\bm{k}$  all details of the bath dropout and its effects can be described by a single frequency independent parameter $\gamma$ describing the effects of inelastic scattering and producing a level broadening $\sim\gamma$ and a finite lifetime $\tau=1/\gamma$.
In this case, the retarded and lesser self-energy become momentum independent and diagonal in Nambu space, {\it i.e.} ${\bm{\Sigma}}_{\bm{k}}^{R}\equiv {\bf I}{{\Sigma}}^{R} $ and  ${\bm{\Sigma}}_{\bm{k}}^{<}\equiv {\bf I}{{\Sigma}}^{<} $
with
\begin{eqnarray}
\nonumber
{\Sigma}^{R}(t_{1},t_{2})&=&-i\gamma\delta(t_{1}-t_{2})/2\,,\\
\Sigma^{<}\left(t_{1},t_{2}\right)&=& i\gamma\int\frac{d\omega}{2\pi}f\left(\omega\right)e^{-i\omega\left(t_{1}-t_{2}\right)}\,.
\label{eq:selfl}
\end{eqnarray}
Here $f(\omega)$ is the Fermi function evaluated at the bath temperature. Consequently, the Dyson equation for the retarded Green function can be easily integrated, being given by ${\bm{G}}_{\bm{k}}^{R}(t,t^{\prime})={\bm{G}}_{\bm{k}}^{R (0)}(t,t^{\prime})e^{-\gamma(t-t^{\prime})/2}$ where ${\bm{G}}_{\bm{k}}^{R (0)}(t,t^{\prime})$ is the retarded Green function in the absence of dissipation. The latter can be computed by solving the following differential equations (in matrix notation in the Nambu spinor basis and setting $\hbar=1$),
\begin{eqnarray}
\label{eq:retev}
\notag
{\bm{G}}_{\bm{\bm{k}}}^{R(0)}\left(t,t\right)&=&-i\bm{I},\\
i\partial_{t}{\bm{G}}_{\bm{\bm{k}}}^{R(0)}\left(t,t^{\prime}\right)&=&\bm{H}_{\bm{k}}(t){\bm{G}}_{\bm{\bm{k}}}^{R(0)}\left(t,t^{\prime}\right),\;\;\;\;\;\; t>t^{\prime},\\
\notag
i\partial_{t^{\prime}}{\bm{G}}_{\bm{\bm{k}}}^{R(0)}\left(t,t^{\prime}\right)&=-&{\bm{G}}_{\bm{\bm{k}}}^{R(0)}\left(t,t^{\prime}\right)\bm{H}_{\bm{k}}\left(t^{\prime}\right),\;\;\;\;\;\; t>t^{\prime}.
\end{eqnarray}
Replacing Eq.~(\ref{eq:selfl}) into Eq.~(\ref{eq:gl}) and assuming a reservoir at zero temperature we obtain
\begin{eqnarray}
\label{eq:glf}
\nonumber
{\bm{G}}_{\bm{k}}^{<}(t,t^{\prime})=-\frac{\gamma}{2\pi}\int_{-\infty}^{t}\!dt_{1}\!\int_{-\infty}^{t^{\prime}}\!dt_{2}\,&&{\bm{ G}}_{ \bm{\bm{k}}}^{R(0)}\left(t,t_{1}\right){\bm{ G}}_{\bm{k}}^{R(0)}(t^{\prime},t_{2})^{\dagger}\\
&&\times\frac{e^{-\gamma(t-t_{1}+t^{\prime}-t_{2})/2}}{t_{1}-t_{2}+i0^{+}}  \,.
\end{eqnarray}
Hence, the time dependence of the order parameter [Eq.~(\ref{eq:deltadef})] can be obtained after computing the lesser Green function [Eq.~(\ref{eq:glf})] for $t^{\prime}=t$. This equal-time lesser Green function $\bm{G}_{\bm{k}}^{<}(t,t)\equiv\bm{G}_{\bm{k}}^{<}(t)$ satisfies the equation of motion
\begin{equation}
\label{eq:gld}
\partial_{t}{\bm{G}}_{\bm{k}}^{<}(t)=-\gamma {\bm{G}}_{\bm{k}}^{<}(t)+\bm{\mathcal{I}}_{\bm{k}}(t)-i\left[\bm{H}_{\bm{k}}(t),{\bm{G}}_{\bm{k}}^{<}(t)\right]\,,
\end{equation}
where
\begin{equation}\label{eq:ik}
\bm{\mathcal{I}}_{\bm{k}}(t)\!=\!\frac{i\gamma}{2\pi}\int_{-\infty}^{t}\!\!dt^{\prime}\left(\frac{{\bm{G}}_{ \bm{k}}^{R (0)}\left(t,t^{\prime}\right)}{t-t^{\prime}-i0^{+}}+\frac{{\bm{G}}_{ \bm{k}}^{R(0)}\left(t,t^{\prime}\right)^{\dagger}}{t-t^{\prime}+i0^{+}}\right)e^{-\frac{\gamma(t-t^{\prime})}{2}}.
\end{equation}
The initial condition for the differential Eq.~(\ref{eq:gld}) is given by the equilibrium value of the lesser Green function
\begin{equation}\label{eq:gl0}
\bm{G}_{\bm{k}}^{<}(0)=\frac{i}{2}\bm{I}-\frac{i}{\pi E_{\bm{k}}}\arctan\left(\frac{2E_{\bm{k}}}{\gamma}\right)\left(\begin{array}{cc}
\xi_{\bm{k}} & -\Delta_{0}\\
-\Delta_{0} & -\xi_{\bm{k}}
\end{array}\right)\,,
\end{equation}
where $E_{\bm{k}}=\sqrt{\xi_{\bm{k}}^2+\Delta_{0} ^2}$, which is time-independent and easily obtained after replacing the equilibrium retarded Green function in Eq.~(\ref{eq:glf}) (see Ref.[\onlinecite{Collado2019}]). 
As a consequence, in the presence of dissipation, the equilibrium order parameter $\Delta_{0}$ is defined, via Eq.~(\ref{eq:deltadef}), by the gap equation
\begin{equation}
\label{eq:gapee}
1=\frac{1}{\pi}\sum_{\bm{k}}\frac{\lambda_{0}}{E_{\bm{k}}}\arctan\left(\frac{2E_{\bm{k}}}{\gamma}\right)\,.
\end{equation}
In the $\gamma\rightarrow  0$ limit Eq.~(\ref{eq:gapee}) becomes the standard BCS gap equation. In the presence of inelastic scattering ($\gamma$) the superconducting order parameter is reduced.  As already mentioned, another important result is that at  equilibrium the present formalism presents a rigorous justification for the Dynes formula for the density of states \cite{Collado2019}. As will be shows explicitly below, this provides a simple way to estimate the $\gamma$ parameter close to equilibrium directly from tunneling experiments \cite{Dynes1978,Saira2012}. 

\subsection{Rabi-Higgs modes in the superconducting response and dissipation effects}
We now present the numerical solution of our model for $\lambda-$driving. For concreteness we shall show simulations for $\omega_d=4\Delta_0$ but qualitative similar behavior is obtained for not too large frequencies above the gap ($\omega_d>2\Delta_0$).

At $t\leq0$ the system is in equilibrium and the order parameter $\Delta_{0}$ is given by Eq.~(\ref{eq:gapee}). At $t>0$ the drive switches on according to Eq.~(\ref{eq:ldt}) and we compute the equal-time lesser Green function via  Eqs.~(\ref{eq:gld}), (\ref{eq:ik}) and (\ref{eq:gl0}) in order to self-consistently determine the superconducting order parameter evolution through Eq.~(\ref{eq:deltadef}).

\subsubsection{Non-linear effects in the undamped dynamics}
We first briefly present the superconducting response in the absence of dissipation as a starting point to compare with those in which the reservoir effects play a role. In this case, the calculation can be made using either the Anderson pseudospin language \cite{Collado2018} or the present formalism by considering the evolution of the equal-time lesser Green function dictated only by the commutator with the  Hamiltonian [Eq.~(\ref{eq:gld}) without the two first terms in the r.h.s.]. 

\begin{figure}[tb]
\includegraphics[width=0.97\columnwidth]{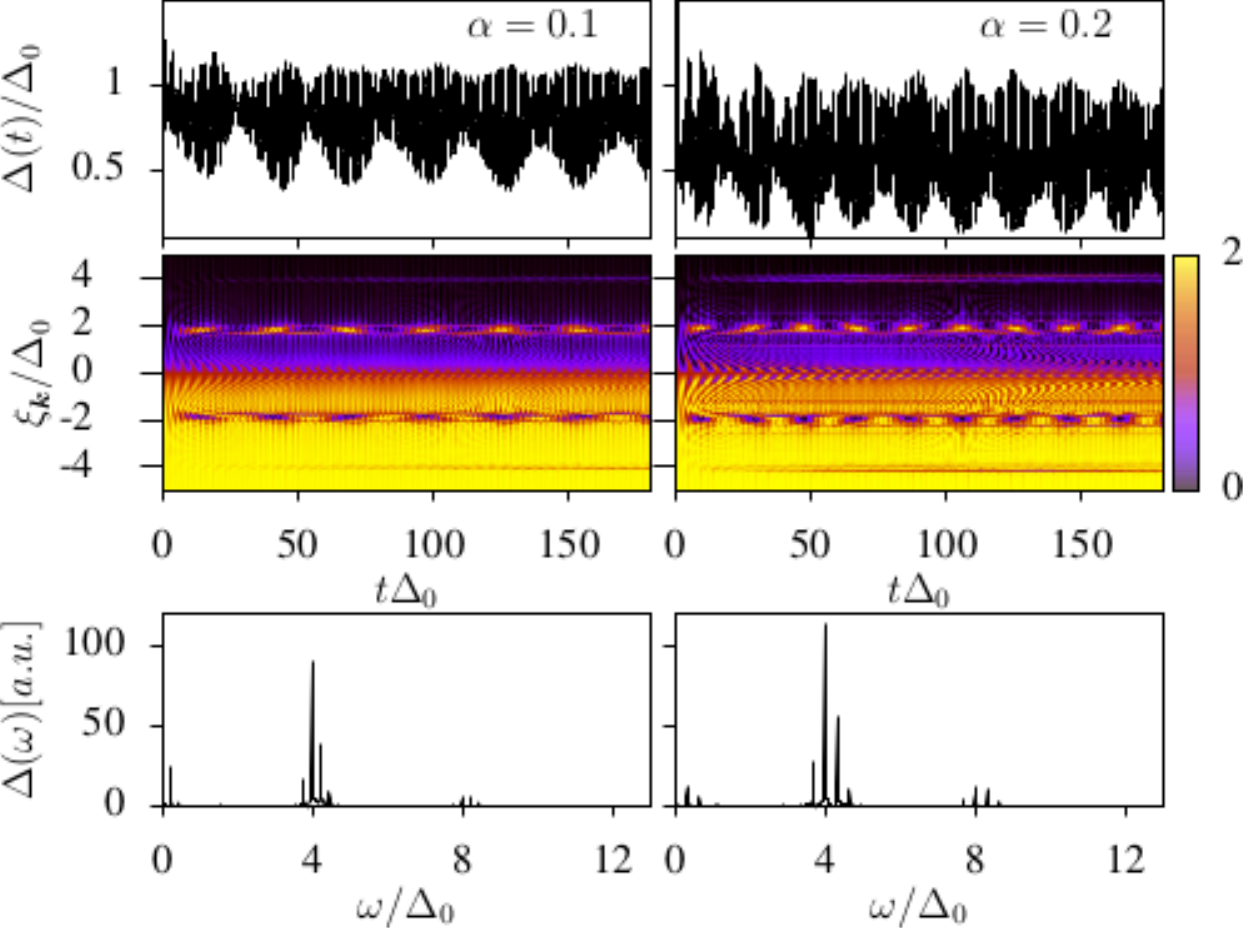}
\caption{(Color online) Superconducting dynamics in the absence of dissipation for $\alpha=0.1$ (left column) and $\alpha=0.2$ (right column) and $\omega_d=4\Delta_0$.
 From top to bottom we show the superconducting order parameter as a function of time, the time-dependent momentum distribution function  $n_{\bm{k}}(t)$ and the fast Fourier transform of $\Delta(t)$. We label the states by the normal state quasiparticle energy $\xi_{\bm{k}}$.} 
\label{fig:fig1}
\end{figure}

The dynamics of $\Delta\left(t\right)$ and the expectation value of the momentum distribution function, 
\begin{equation}
n_{\bm{k}}(t)=\sum_{\sigma}\left\langle c_{\bm{k}\sigma}^{\dagger}(t)c_{\bm{k}\sigma}(t)\right\rangle = 1-i \left[\bm{G}_{\bm{k}}^{<}(t)_{11}-\bm{G}_{\bm{k}}^{<}(t)_{22}\right]
\end{equation}
are shown in Fig.~\ref{fig:fig1} for two different values of the perturbation amplitude $\alpha$.
It is apparent from the figure that the order parameter oscillates with two fundamental frequencies and, after a short transient, averages to a smaller value respect to equilibrium. The drive frequency corresponds to a fast oscillation that can not be resolved on the scale of the figure and leads to the filled black regions of the gap dynamic. In addition, the amplitude of the gap shows the Rabi-Higgs oscillations with a frequency that increases approximately linearly with increasing $\alpha$. Indeed, 
\begin{eqnarray}
  \label{eq:wr}
\frac{\Omega_R}{\Delta_0} \approx A(\omega_d/\Delta_0) \alpha
\end{eqnarray}
with $A(\omega_d/\Delta_0)\sim 2.0$ for  $\omega_d/\Delta_0=4$ [see Eq. (29) in Ref.~\cite{Collado2018} for an analytic approximation].

The  Rabi-Higgs mode is associated with a periodic inversion of the population of the quasiparticles resonant with the drive. For $\omega_d=4\Delta_0$, this is visible in the momentum distribution function $n_{\bm{k}}(t)$  as a narrow time dependent structure at quasiparticle energy  $\xi_{\bm{k}}\approx\pm 2\Delta_{0}$   (Fig.~\ref{fig:fig1}, middle panel) with a frequency that  matches the Rabi-Higgs period of $\Delta(t)$. Notice that the inversion of colors along the anomaly represent a cyclic inversion of population of the resonant quasiparticles.  
Such time-dependent anomaly represent a clear hallmark of the Rabi-Higgs mode and opens the possibility  to detect it through spectroscopies as we shall demonstrate in the next section. 

Another non-linear effect is the generation of a second-harmonic. In the case of an electromagnetic drive, second harmonic generation is not allowed \cite{Cea2018}. This follows from the general fact that the current response to the vector potential $A$ is $J\sim \rho_s(A) A$ where $\rho_s(A)$ is the superfluid stiffness. In the absence of a steady state current, the free energy and  $\rho_s(A)$ are even in $A$, so the lowest non-linear contribution to $J$ is order $A^3$, i.e. third harmonic generation \cite{Cea2018}. In contrast, not such symmetry exist in our case as for $\lambda>0$ terms odd in $\delta \lambda$ are allowed in the free energy and second harmonic generation is allowed. Indeed, as can be seen in the bottom panel of Fig.~\ref{fig:fig1} the superconducting response not only contains the driving frequency $\omega_d=4\Delta_{0}$ but also $2\omega_d=8\Delta_{0}$. 
This can also been seen in the middle panel of Fig.~\ref{fig:fig1}, where a less intense Rabi-Higgs mode is developed associated with the second harmonic response, i.e. quasiparticles with $\xi_{\bm{k}}\approx\pm 4\Delta_{0}$ also show a narrow time dependent anomaly in the population but with a much smaller Rabi frequency. 

\begin{figure}[t]
\includegraphics[width=0.95\columnwidth]{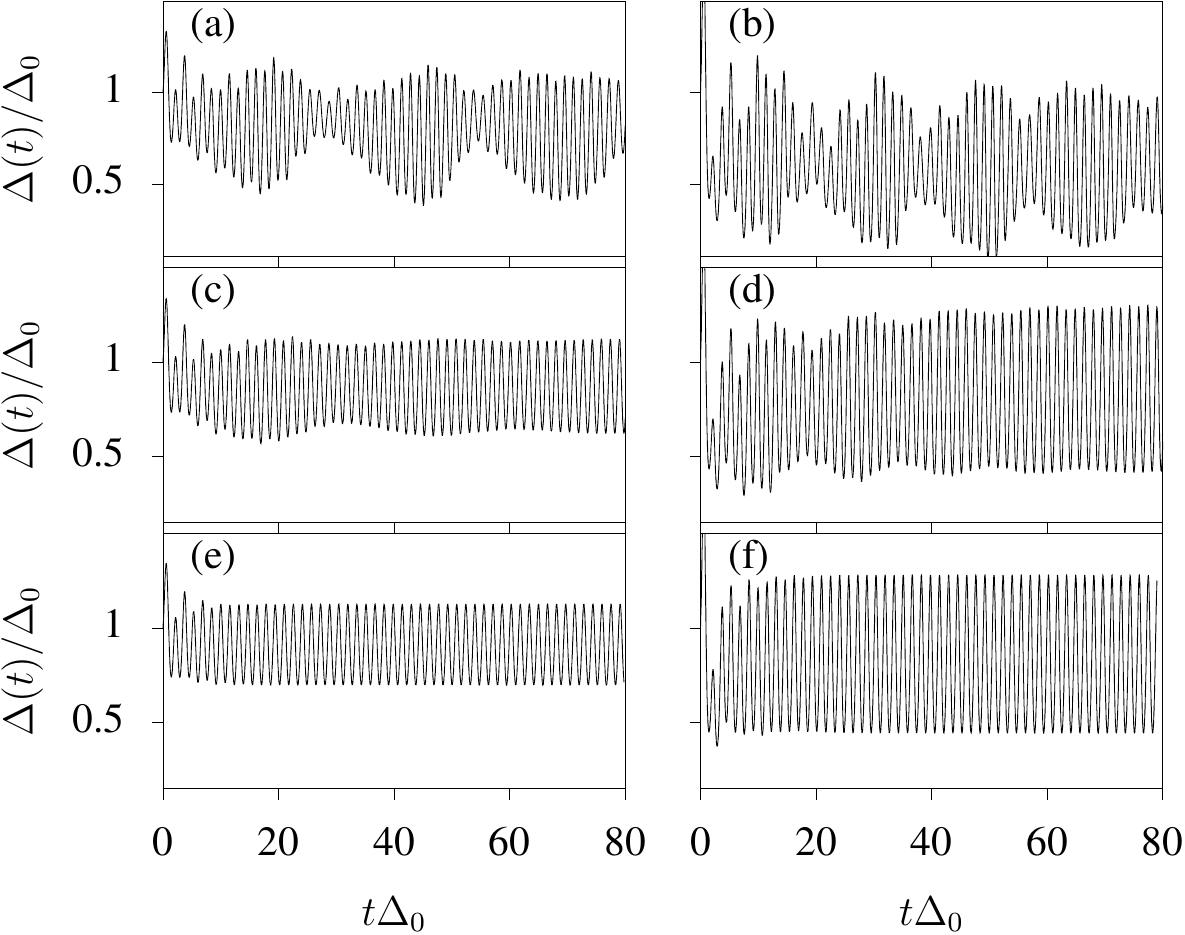}
\caption{Time dependence of the superconducting order parameter in the presence of dissipation for $\alpha=0.1$ (left column) and $\alpha=0.2$ (right column). From top to bottom we use $\gamma=0$ (without dissipation effects), $\gamma=0.05\Delta_{0}$ and $\gamma=0.2\Delta_{0}$, respectively.} 
\label{fig:fig2}
\end{figure}

\subsubsection{Dissipative dynamics}
The driven superconducting response ($\Delta(t)$) in the presence of a bath is shown in Fig.~\ref{fig:fig2} for different values of the bath parameter $\gamma$ and the perturbation amplitude $\alpha$. As in non-linear optics, in the presence of dissipation the Rabi oscillation is a transient effect. For  sufficiently long times a steady state is achieved where only the drive frequency is present.

 As $\gamma$ is increased (from top to bottom in Fig.~\ref{fig:fig2}) the system reaches a steady state more rapidly with a vanishing of Rabi-Higgs oscillations. By increasing $\alpha$, for a fixed value of $\gamma$, $\Omega_R$ increases [cf. Eq.~(\ref{eq:wr})] and several Rabi-Higgs oscillations are visible before they disappear as a consequence of relaxation (see for example Fig.~\ref{fig:fig2} (c,d)). Thus, in order to detect the Rabi-Higgs modes experimentally it is necessary to ensure that $\Omega_R \gtrsim\gamma$.  In fact, for the $\gamma$ values used here, the slower Rabi-Higgs mode, associated with the second harmonic generation, is not visible.

\begin{figure}[t]
\includegraphics[width=0.95\columnwidth]{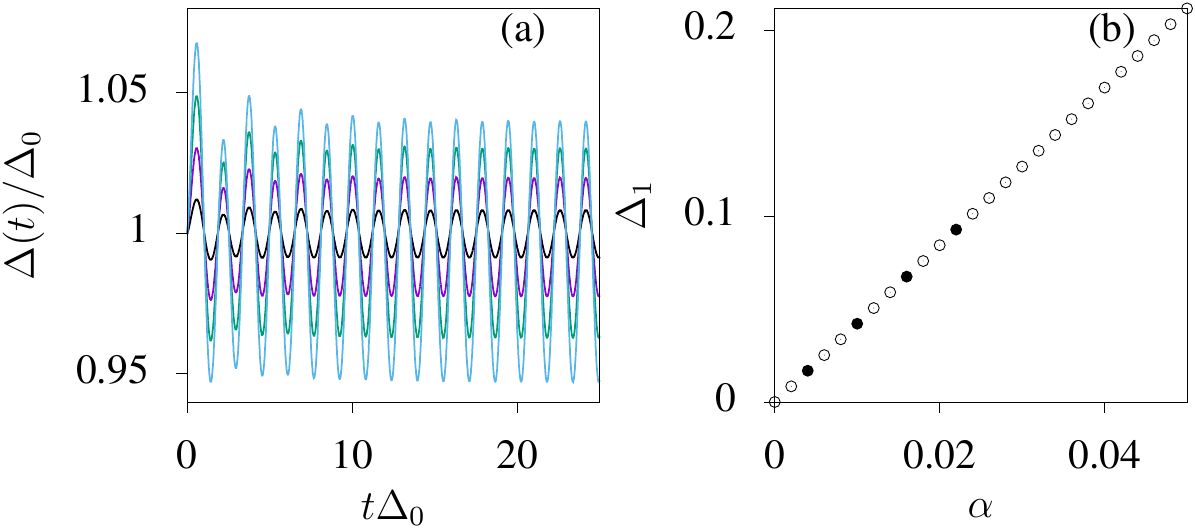}
\caption{(Color online) (a) Time dependence of the order parameter for $\gamma=0.2\Delta_{0}$ and $\alpha=0.004, 0.01, 0.016, 0.022$. (b) Amplitude of the oscillation in the order parameter $\Delta_1$ as a function of the strength of the drive $\alpha$. The filled points correspond with the curves of panel (a).} 
\label{fig:fig21}
\end{figure}

Conversely, for  small $\alpha$ and strong dissipation it is possible to get $\Omega_R\lesssim\gamma$ and only the oscillations synchronous with the drive are visible as expected from  linear response theory (bottom panels of Fig.~\ref{fig:fig2} and Fig.~\ref{fig:fig21}). Indeed, in this regime, the amplitude of the oscillation in the order parameter increases linearly with $\alpha$ as  shown in Fig.~\ref{fig:fig21}. However, it is important to note that by increasing $\alpha$, and after a very fast transient, the superconducting gap decreases in average which constitutes the first nonlinear effect arising in the dynamics. As a conclusion, in the presence of dissipation, we can distinguish two different regimes: one in which the linear response theory is applicable even at short times ($\Omega_R\lesssim\gamma$) and the other corresponding with a transient strong nonlinear behavior in which Rabi oscillations can be observed ($\Omega_R\gtrsim\gamma$).

We now discuss in more detail the nonlinear regime, in particular the possibility of detection of the Rabi-Higgs mode in experiments. Previously, we have demonstrated that the Rabi-Higgs mode is associated with oscillations in the occupation values $n_{\bm{k}}$ (see Fig.~\ref{fig:fig1}). This charge fluctuations provide an efficient manner to detect the existence of this nonlinear mode with standard experimental techniques as we shall show in the next section. 

Since for the simple electronic structure we are taking, $n_{\bm{k}}$ depends on ${\bm{k}}$ only through its distance from the Fermi surface ($\xi_{\bm{k}}$) it is useful to introduce the distribution function $ n(\xi_{\bm{k}})\equiv n_{\bm{k}}$. When convenient, in the following we will drop the momentum dependence and refer to $ n(\xi)$ loosely as the momentum distribution function keeping in mind the above equivalence. 

\begin{figure}[tb]
\includegraphics[width=0.95\columnwidth]{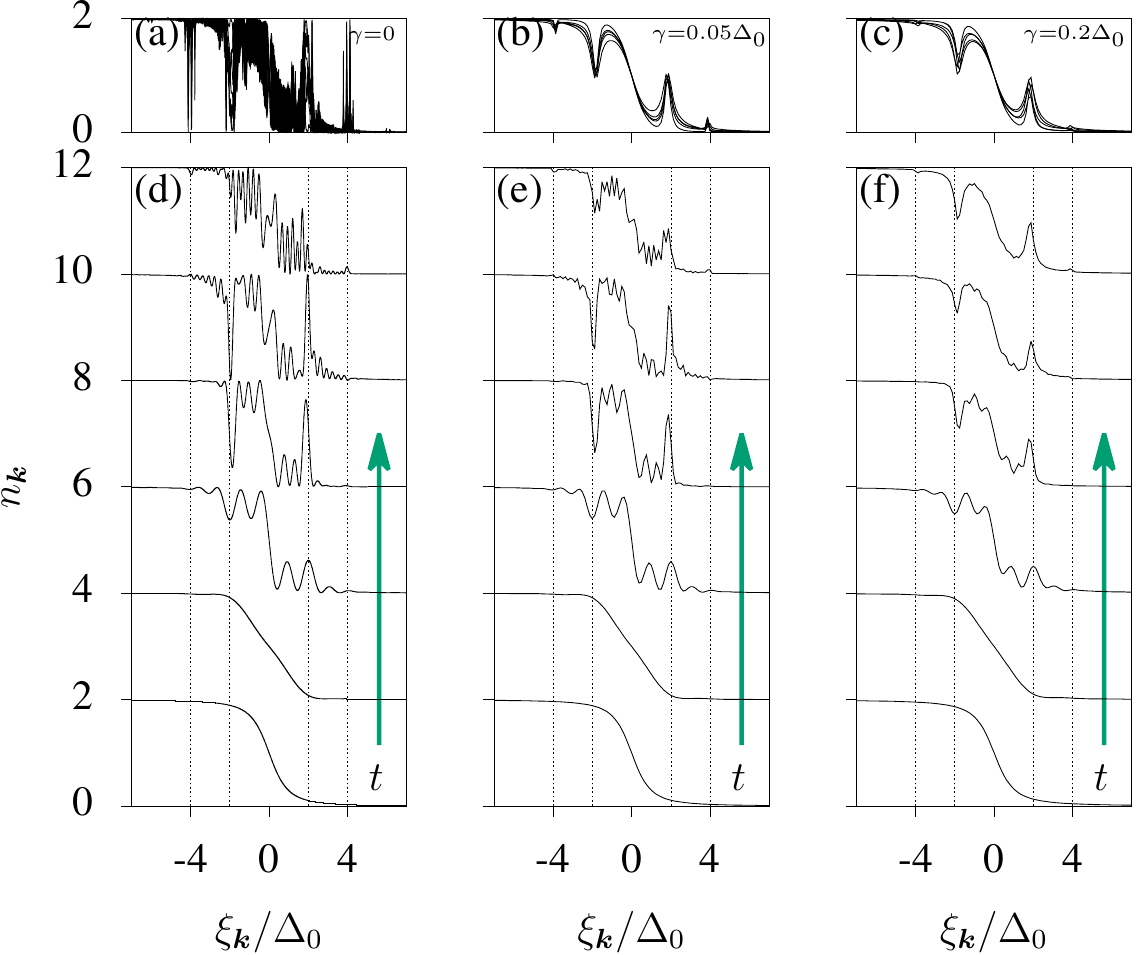}
\caption{$n_{\bm{k}}$ distribution as a function of time for $\alpha=0.2$ and $\gamma=0$ (left column),  $\gamma=0.05\Delta_{0}$ (middle column) and $\gamma=0.2\Delta_{0}$ (right column). In panels (d)-(f), the time increases from bottom to top as follows $t=0$ (equilibrium), $t=0.1\tau_R$, $t=0.25\tau_R$,  $t=0.5\tau_R$,  $t=0.75\tau_R$ and  $t=\tau_R$ (Rabi-Higgs period $\tau_R \Delta_0=15$). For clarity, the zero of each curve was displaced vertically for a factor 2. The oscillating $n_{\bm{k}}$ distributions in the steady state are shown in panels (a)-(c) at different times within a drive period.} 
\label{fig:fig3}
\end{figure}

In Fig.~\ref{fig:fig3} we show the time-dependent $n_{\bm{k}}$ distribution taking into account dissipative effects (middle and right column) in comparison with the dissipationless counterpart (left column) at different times within the first Rabi cycle (panels (d)-(f)) and in the steady state (panels (a)-(c)). We have used the same $\gamma$ and $\alpha$ parameters as in the right column of Fig.~\ref{fig:fig2}. 

For $\gamma=0$ the Rabi oscillations can be observed as oscillations in the occupation value $n_{\bm{k}}$ for $\xi_{\bm{k}}\approx\pm 2\Delta_{0}$, which we will refer to as  $n({\pm}\xi_R)$. Small peaks also appear for $\xi_{\bm{k}}\approx\pm 4\Delta_{0}$, corresponding to the second-harmonic  slower Rabi-Higgs mode that starts to develop in the temporal window used in Fig.~\ref{fig:fig3}(d). At long-times, the two Rabi-Higgs modes can be seen as is shown in Fig.~\ref{fig:fig3}(a). 

In the presence of weak dissipation ($\gamma=0.05\Delta_{0}$), the transient dynamic only shows oscillations of  $n(\pm\xi_R)$, corresponding to the fastest Rabi-Higgs mode. Finally, 
for strong dissipation ($\gamma=0.2\Delta_{0}$), a full Rabi cycle cannot be completed since the relaxation takes place in a very short time. As a consequence $n(\xi_R)$ ($n(-\xi_R)$) increases (decreases) during a short period of time and rapidly saturates without exhibiting Rabi oscillations.

The  $n_{\bm{k}}$ distribution in the steady-state in the presence of dissipation are shown in Fig.~\ref{fig:fig3}(b) and Fig.~\ref{fig:fig3}(c) for different times within a period of the drive. In this case, $n_{\bm{k}}$ shows a peak at $\xi_{\bm{k}}\approx\pm 2\Delta_{0}$ such that $n(\pm\xi_R)\approx 1$ all the time. Taking into account spin this correspond to half-filled single particle states which, as shown below, corresponds to saturation of the population imbalance due to strong drive. Thus, ultimately the fate of the superconducting dynamics is a gap oscillation with the drive frequency (see Fig.~\ref{fig:fig2} (e,f)) and a stationary population unbalance at the quasiparticle energy $\xi_{R}$. 

\begin{figure}[tb]
\includegraphics[width=0.95\columnwidth]{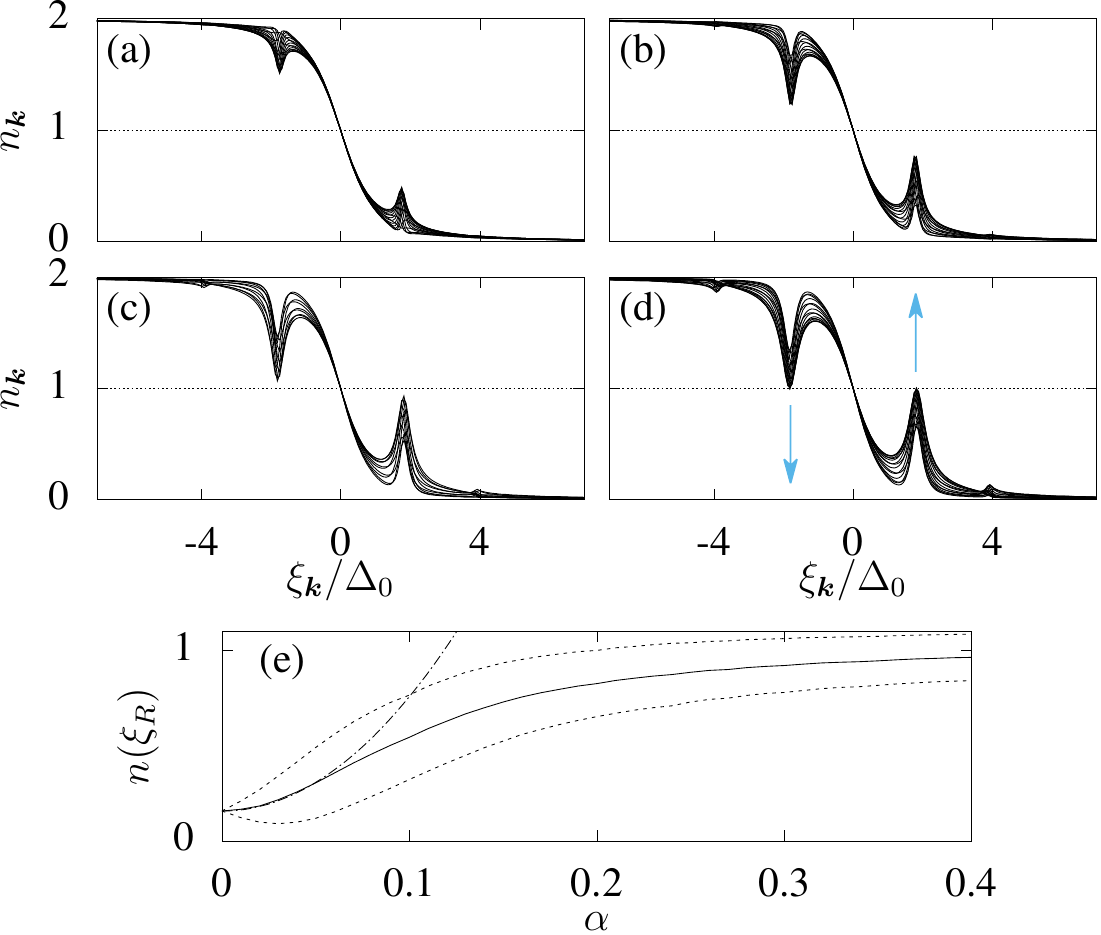}
\caption{Steady-state $n_{\bm{k}}$ distributions for $\gamma=0.2\Delta_0$ and $\alpha=0.05$ (a), $\alpha=0.1$ (b), $\alpha=0.15$ (c) and $\alpha=0.2$ (d). In panel (e) the dashed and solid lines represent the (minimum, maximum) and average value of $n(\xi_R)$, respectively, as a function of $\alpha$. Notice that  $n(\xi_R)$ is non zero even for $\alpha=0$ as an excess of occupation is inherent to the BCS ground state. The dot-dashed line is a quadratic fitting for the average of excited population for small $\alpha$ values. The population inversion starts when $n(\xi_R)$ exceeds the horizontal line $n_{\bm{k}}=1$ which is represented with arrows in the panel (d). As a consequence of the presence of dissipation, in the steady state regime this does not occur not matter how strong the perturbation is.} 
\label{fig:fig31}
\end{figure}

The population unbalance changes considerably as a function of the amplitude of the perturbation as can be appreciated in Fig.~\ref{fig:fig31}. Because the Rabi oscillation is only a transient effect, a full population inversion (e.g. a sign change of $n(\xi_R)-n(-\xi_R)$) is not possible in the steady state. As a consequence, the excited state population can not exceed the half-filled value (counting spin) $n(\xi_R)\approx1$ by considering a bath in the wide-band limit \cite{Ferron2012}.

We have extracted the $n_{\bm{k}}$ distribution in the steady state for several values of $\alpha$ with $\gamma=0.2\Delta_{0}$. In accordance with the above statement, the value of $n(\xi_R)$ ($n(-\xi_R)$) in the steady state, increases (decreases) with $\alpha$ and saturates to $1$ in the limit of large $\alpha$ as it is shown in Fig.~\ref{fig:fig31}. Notice that the population imbalance is quadratic in $\alpha$ so it vanishes in the linear response regime which follows also from the fact that, by symmetry, it should be even in $\alpha$.  

This saturation of the excited population is a common effect in nonlinear quantum optics. There, one usually considers a two-level system in the presence of a driving force and damping effects via a phenomenological parameter in the Bloch equation of motion. For a drive frequency in resonance with the two level system, the excited population in the steady state increases with the amplitude of the perturbation and in the limit of large intensity, the largest possible excited population is equal to the ground state population \cite{Steck2019}. Again, the difference in the present context is that the phenomena does not corresponds to a single two-level system but a collection of two-level systems  interacting through the self-consistency that determines the superconducting order parameter. 

\section{Theoretical modelling of tr-ARPES and Tunneling experiments: Spectral fingerprints of Rabi-Higgs mode}
Our main aim in this section is to identify some spectral fingerprints of the Rabi-Higgs mode, and related non-linear phenomena, that could be experimentally detected in the presence of  dissipation. 
Clearly, an emergent technique to detect time dependent phenomena is tr-ARPES and so we discuss such a case first. Yet,  in addition, we shall demonstrate that also time resolved tunneling experiments could be useful to detect the non-linear mode discussed above.

\begin{figure}[tb]
\includegraphics[width=0.95\columnwidth]{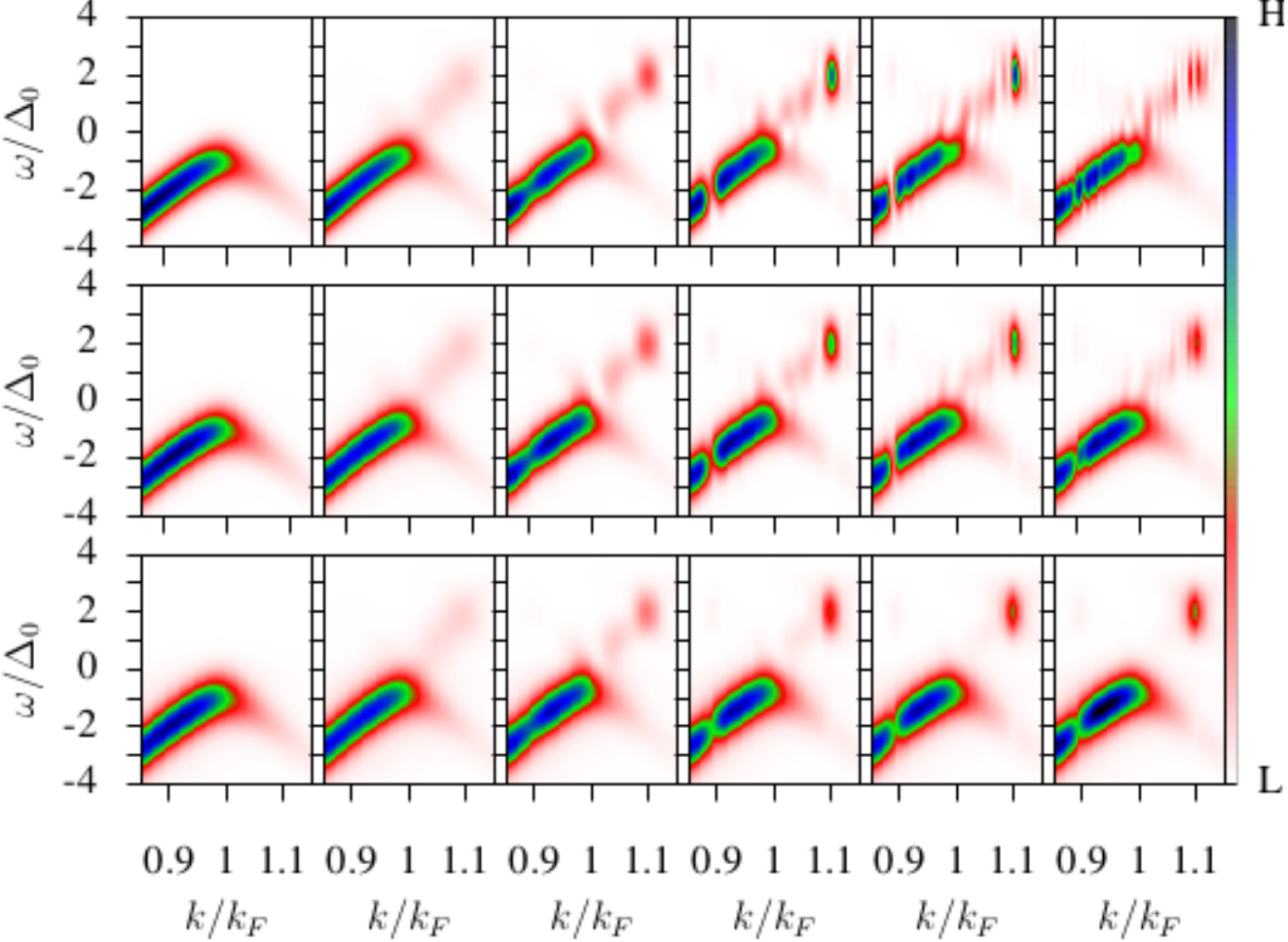}
\caption{(Color online) tr-ARPES intensity for $\alpha=0.2$ at several photo-emission times inside the first period $\tau_R$ of Rabi-Higgs mode, from left to right, $t_0=0$ (equilibrium), $t_0=0.1\tau_R$, $t_0=0.25\tau_R$,  $t_0=0.5\tau_R$,  $t_0=0.75\tau_R$ and  $t_0=\tau_R$. From top to bottom we show results for $\gamma=0$ (without dissipation), $\gamma=0.05\Delta_{0}$ and $\gamma=0.2\Delta_{0}$ as in Fig.~\ref{fig:fig2}.} 
\label{fig:fig4}
\end{figure}

In all the calculations discussed below, we obtain the spectral signals in terms of the lesser Green function Eq.~(\ref{eq:glf}) as a function of $\alpha$ and $\gamma$. 
\subsection{tr-ARPES}
In our setting the  $\lambda-$driving is turned on at time $t=0$ and the photoemission process is induced by a wave packet of photons centered at time $t_0$ and with central energy $\hbar \omega_{\bm q}$ larger than the work function of the solid $W$. For simplicity we use a Gaussian shape for this probe pulse, $s(t)=\exp(-(t-t_{0})^{2}/2\sigma^{2})/(\sigma\sqrt{2\pi})$ with standard deviation $\sigma$. 

In a tr-ARPES experiment the momentum of the outgoing electrons ${\bm k}_e$ is measured. Energy conservation determines the excitation energy left in the system after the photoemission process,
$\hbar \omega=\hbar\omega_{\bm q}-(\hbar k_e)^2/(2m_e)-W$ and momentum conservation yields information on the momentum of the excitations ${\bm{k}}$. The momentum resolved photocurrent in the detector at time $t$ is due to all electrons photoemitted before that time and is determined by
\cite{Freericks2009},
\begin{equation}
\label{eq:arpesint}
I_{\bm{k}}\left(\omega,t\right)=\mathrm{Im}\!\!\int_{-\infty}^{t}\!\!dt_{1}\!\!\int_{-\infty}^{t}\!\!dt_{2}s\left(t_{1}\right)s\left(t_{2}\right)e^{i\omega\left(t_{1}-t_{2}\right)}\bm{G}_{\bm{k}}^{<}\left(t_{1},t_{2}\right)_{11}\,.
\end{equation}

In order to probe the Rabi-Higgs modes in real time in the following we use 
different probe times $t_{0}$ and take the integration limit from a lower cutoff time $t_l=t_0-5\sigma$ to $t=t_0+5\sigma$. Fig.~\ref{fig:fig4} shows the tr-ARPES intensity  with and without dissipation for different $t_0$ during the first period $\tau_R$ associated with the Rabi-Higgs mode. For simplicity, we consider a parabolic band for electrons $\xi_{\bm{k}}\propto (k^{2}-k_F^{2})$ and use a probe pulse with a standard deviation $\sigma=0.1\tau_R$. From the photo-emission signal at equilibrium (left column in Fig.~\ref{fig:fig4}) there is an almost imperceptible broadening of the  spectral line as $\gamma$ is increased, which leads to a decrease of superconducting order parameter according to Eq.~(\ref{eq:gapee}).

The presence of Rabi-Higgs oscillations are clearly visible in the top and middle panels of Fig.~\ref{fig:fig4} via a spectral weight around $\omega=\pm2\Delta_{0}$ that increases  (decreases) above (below) the Fermi energy in the first half-period and has the opposite behavior in the second 
half-period of the Rabi-Higgs mode. One can visualize the process as an excitation of quasiparticles from the lower quasiparticle branch to the higher quasiparticle branch in the first half-cycle followed by a deexcitation in the second half-cycle or equivalently as a stimulated absorption phase followed by a stimulated emission phase. 

If $\gamma$ is large enough, Rabi oscillations are not visible and only the spectral weight corresponding with the steady state is observed after a fast transient (bottom panel of Fig.~\ref{fig:fig4}). 

Clearly the above dynamics is the photoemission image of the momentum distribution function imbalance discussed above. Indeed, integrating in frequency 
\begin{eqnarray}
  \label{eq:sumrule} \int_{-\infty}^{\infty}\frac{d\omega}{2\pi}I_{\bm{k}}\left(\omega,t\right)&=&\mathrm{Im}\!\!\int_{-\infty}^{t}\!\!dt^{\prime}s\left(t^{\prime}\right)^2\bm{G}_{\bm{k}}^{<}\left(t^{\prime},t^{\prime}\right)_{11}\nonumber\\
&=&\frac12\int_{-\infty}^{t}\!\!dt^{\prime}s\left(t^{\prime}\right)^2 n_{\bm{k}}(t')
\end{eqnarray}
which is clearly a moving average of  $n_{\bm{k}}(t)$. For example, in the large $\gamma$ case the 
unbalance over time in the tr-ARPES intensity for $\omega=\pm2\Delta_{0}$ matches the $n_{\bm{k}}$ distribution in the steady state shown in Fig.~\ref{fig:fig31}(d).

These results establish  time-resolved photo-emission experiments as a tool to investigate steady state non-linearities, like population imbalance, and dynamic non-linearities like the Rabi-Higgs mode, in driven superconductors. 

\subsubsection{Floquet Analysis}

As already mentioned for $t\gtrsim\tau=1/\gamma$ one reaches a steady state and the superconducting gap oscillates only with the drive frequency. Therefore the mean-field Hamiltonian  Eq.~(\ref{eq:hmf}) is time periodic at long times and we can use Floquet theorem to analyze the spectrum. This theorem guarantees the existence of a set of solutions of the time-dependent Schrodinger equation of the form
\begin{equation}
\left|\psi_{\nu}\left(t\right)\right\rangle=\exp(-i\varepsilon_{\nu}t/\hbar)\left|\phi_{\nu}\left(t\right)\right\rangle
\end{equation}
where $\left|\phi_{\nu}\left(t\right)\right\rangle$ has the same periodicity of the Hamiltonian Eq.~(\ref{eq:hkt})~\cite{Shirley65,Sambe73}. The Floquet states $\left|\phi_{\nu}\left(t\right)\right\rangle$ are the solutions of 
\begin{equation}
\mathcal{H}_{F}\left|\phi_{\nu}\left(t\right)\right\rangle=\varepsilon_{\nu}\left|\phi_{\nu}\left(t\right)\right\rangle
\end{equation}
where $\mathcal{H}_{F}=H_{MF}-i\hbar\partial/\partial t$ is the Floquet Hamiltonian and $\varepsilon_{\nu}$ is the quasienergy~\cite{Kohler2005,Grifoni98}. In the Floquet basis this Hamiltonian becomes a time-independent infinite matrix operator. Since we are interested in the low energy spectra we restrict the Floquet Hamiltonian to a large enough but finite subspace containing many multiphoton processes (finite number of replicas)~\cite{Kohler2005,Grifoni98}. 

For $\gamma=0.2\Delta_0$ and $\alpha=0.2$ the steady state condition is reached already for $t\sim 0.75 \tau_R= 11.25/\Delta_0$. The quasi-energy spectrum in the long time limit is well described by  $\Delta(t)=\bar{\Delta}+\Delta_1 \cos(\omega_d t)$ with $\bar{\Delta}=0.86\Delta_0$ and $\Delta_1=0.42\Delta_0$  [c.f. Fig.~\ref{fig:fig2}(f)].

\begin{figure}[tb]
\includegraphics[width=0.95\columnwidth]{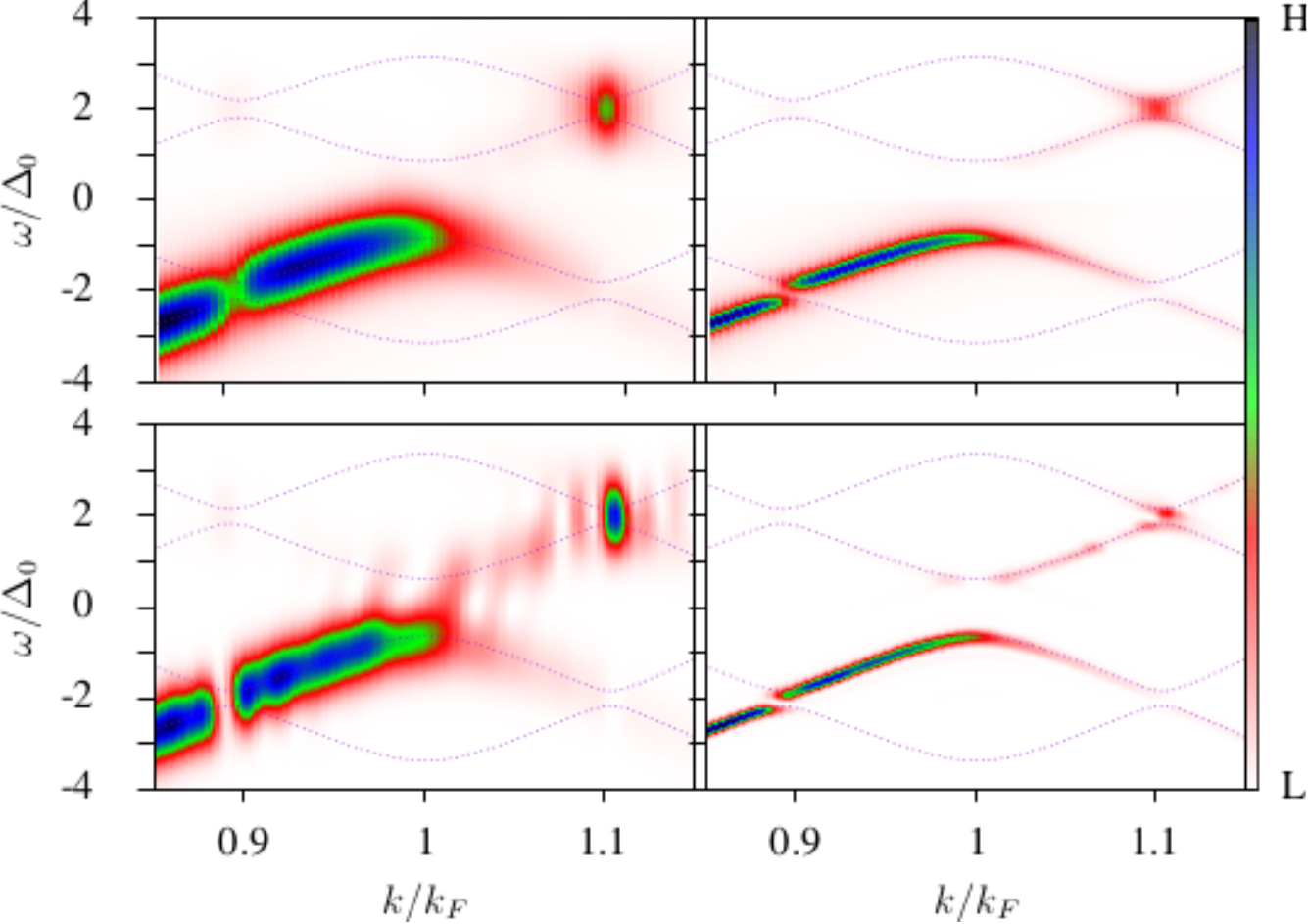}
\caption{(Color online) tr-ARPES intensity for drive strength $\alpha=0.2$, $t_0=0.75\tau_R$ and $\gamma=0.2\Delta_0$ (upper panel) and $\gamma=0$ (lower panel). We use a probe duration of $\sigma=0.1\tau_R$ (left column) and $\sigma=0.6\tau_R$ (right column). The dashed lines represent the quasi-energy spectrum assuming a periodic time dependence $\Delta(t)$ as defined in the main text.} 
\label{fig:fig41}
\end{figure}

In the upper panels of Fig~\ref{fig:fig41} we compare the tr-ARPES with the Floquet spectrum in the steady state. To analyze the details in the right panels we sacrifice temporal resolution to gain energy resolution by using a wider probe pulse. We see that the tr-ARPES signal nicely matches the Floquet spectrum, depicted by the dashed line. Thus the tr-ARPES signal essentially probes the occupied parts of the Floquet band structure.

Surprisingly, also in the transient dynamics the tr-ARPES intensity fits very well with a Floquet spectrum that is obtained from an effective $\Delta(t)$ with a monochromatic dependence (see lower panel of Fig.~\ref{fig:fig41}), even thought in this regime the superconducting gap shows several incommensurate frequencies (Rabi-Higgs and drive frequency) and Floquet theorem is not strictly applicable. For $\gamma=0$, we compute the Hamiltonian spectrum for a $\Delta(t)=0.63\Delta_0+0.35\Delta_0\cos(\omega_d t)$. Thus the transient response shown in Fig.~\ref{fig:fig4} and the lower panels in Fig.~\ref{fig:fig41} can be seen as a  time-dependent change in the occupancy of the Floquet band structure. Clearly, the reason this analysis works in the Rabi-Higgs oscillation regime is the large separation of time scales between the slow Rabi-Higgs dynamics and the fast drive oscillations \cite{Lucila2018}. 

\subsection{Time-resolved tunneling experiment}
We now discuss a possible setup to detect the Rabi-Higgs mode via tunneling measurements. In the last decades there have been efforts to add temporal resolution to scanning tunneling spectroscopy (STM) techniques and now sub-picosenconds resolution can be reached in experimental setups~\cite{Morgenstern1609,Nunes1029,Mikio2004}. In principle one can radiate a sample while performing the measurement  making STM a feasible time resolved technique at the same level that pump-probe measurements \cite{Mikio2004}. However, since the spatial resolution is not a requirement to reveal this nonlinear mode, a planar junction would probably be a more stable setting. An antenna could be used to couple the radiation with the superconductors as is done in detectors for photons with energy matching the superconducting gap \cite{Zmuidzinas1992,Leridon1997}. In what follows, we will refer to STM signal but our formalism applies also for the case of a planar junction.

To obtain the STM signal, we now consider a tip located close above the out of equilibrium superconductor and tunnel-coupled locally to the $\bm{k}$ states in the system. The entire problem can be described by a generic Hamiltonian $H=H_{\mathrm{MF}}+H_{\mathrm{T}}+H_{\mathrm{TS}}$ where we have added the tip (or metallic contact) Hamiltonian 
\begin{equation}
H_{\mathrm{T}}=\sum_{\bm{p}}(\epsilon_{\bm{p}}+eV)a_{\bm{p}}^{\dagger}a_{\bm{p}}^{}\,,
\end{equation}
and the coupling between both subsystem via the tunneling Hamiltonian
\begin{equation}
H_{\mathrm{TS}}=\sum_{\bm{k},\bm{p}}\left(T_{\bm{k}\bm{p}}\,c_{\bm{k}}^{\dagger}a_{\bm{p}}^{}+h.c.\right).
\end{equation}
The tip or the metallic contact is connected to an external voltage source with energy $eV$, $a_{\bm{p}}$ ($a_{\bm{p}}^{\dagger})$ destroys (creates) an electron with momentum $\bm{p}$ and energy $\epsilon_{\bm{p}}+eV$ in the tip and $T_{\bm{k}\bm{p}}$ is the strength of the tunneling. The electron current operator is given by
\begin{equation}
\hat{I}=ie\left[\hat{N},\hat{H}\right]=ie\left(\hat{L}^{\dagger}-\hat{L} \right)\,,
\end{equation}
where $e$ is the electron charge, $\hat{N}=\sum_{\bm{p}}a_{\bm{p}}^{\dagger}a_{\bm{p}}$ is the number operator and $\hat{L}=\sum_{\bm{k},\bm{p}}T_{\bm{k}\bm{p}}c_{\bm{k}}^{\dagger}a_{\bm{p}}^{}$. Assuming a weak coupling between tip and the superconductor we calculate the current through the tip in the linear response limit as
\begin{equation}
\label{eq:itd}
I(t)=ie\int_{-\infty}^{\infty}\theta\left(t-t^{\prime}\right)\left\langle \left[\hat{I}(t),H_{TS}(t^{\prime})\right]\right\rangle_0 
\end{equation}
where $\left\langle ... \right\rangle_0 $ denotes the expectation value at zero order in the tunneling Hamiltonian. Notice that we have not consider the spin degree of freedom so far as the current is a spin conserved quantity (not spin-flip are allowed in the tunneling process). This will be included in the final expression of the time-dependent current as a factor $2$. From Eq.~(\ref{eq:itd}) we obtain 
\begin{widetext}
\begin{equation}
\label{eq:itf}
I(t)=2eT^{2}\int_{-\infty}^{t}dt^{\prime}\sum_{\bm{k,p}}\left(\left[e^{-i\epsilon_{\bm{p}}\left(t-t^{\prime}\right)}\left\langle c_{\bm{k}}^{\dagger}(t)c_{\bm{k}}^{}(t^{\prime})\right\rangle \Theta\left(\epsilon_{\bm{p}}-eV\right)-e^{i\epsilon_{\bm{p}}\left(t-t^{\prime}\right)}\left\langle c_{\bm{k}}^{}(t)c_{\bm{k}}^{\dagger}(t^{\prime})\right\rangle \Theta\left(-\epsilon_{\bm{p}}+eV\right)\right]+\left(t\leftrightarrow t^{\prime}\right)\right)\,,
\end{equation}
\end{widetext}
where we have assumed a momentum-independent tunneling coupling $T_{\bm{k}\bm{p}}\equiv T$ and a zero-temperature Fermi distribution for the electrons on the tip via the Heaviside step function $\Theta(x)$. Taking the derivative of Eq.~(\ref{eq:itf}) with respect to the voltage $V$ and using Eq.~(\ref{eq:greendeff}), the time-dependent differential conductance can be written as
\begin{equation}
\label{eq:conductance}
G(t)=\frac{dI}{dV}\propto \mathrm{Im}\sum_{\bm{k}}\int_{-\infty}^{t}dt^{\prime}\mathrm{Tr}\left[{\bm{G}}_{\bm{k}}^{<}(t,t^{\prime})\right]e^{-ieV\left(t-t^{\prime}\right)}
\end{equation}
where $\mathrm{Tr}$ represents the trace in Nambu space.
In the absence of the drive (that is, at equilibrium) the differential conductance becomes time-independent, being proportional to the well-know phenomenological Dynes density of states
\begin{equation}
\label{eq:dos}
G\propto\rho(eV)=\rho_{0}\, \mathrm{Re} \left [ \frac{eV+i \Gamma}{\sqrt{(eV+i \Gamma)^2-\Delta_{0}^2}}\right ]\,,
\end{equation}
where the Dynes parameter $\Gamma=\gamma/2$, $\Delta_{0}$ is the equilibrium order parameter and $\rho_{0}$ is the normal phase density of states. It is worth recalling that the Dynes formula was originally introduced phenomenologicaly \cite{Dynes1978}. There have been several theoretical proposals to provide a formal justification including Eliashberg physics \cite{Mikhailovsky1991}, inelastic tunneling \cite{Pekola2010} and magnetic impurities \cite{Hlubina2016,Hlubina2018}. In our previous work we have shown that the coupling with a bath \cite{Collado2019} provides a mechanism to justify Dynes formula and allows to link directly equilibrium tunneling with the $\gamma$ parameter. Of course since we are not providing a microscopic theory of the bath our justification is still semi-phenomenological.

\begin{figure}[tb]
\includegraphics[width=0.95\columnwidth]{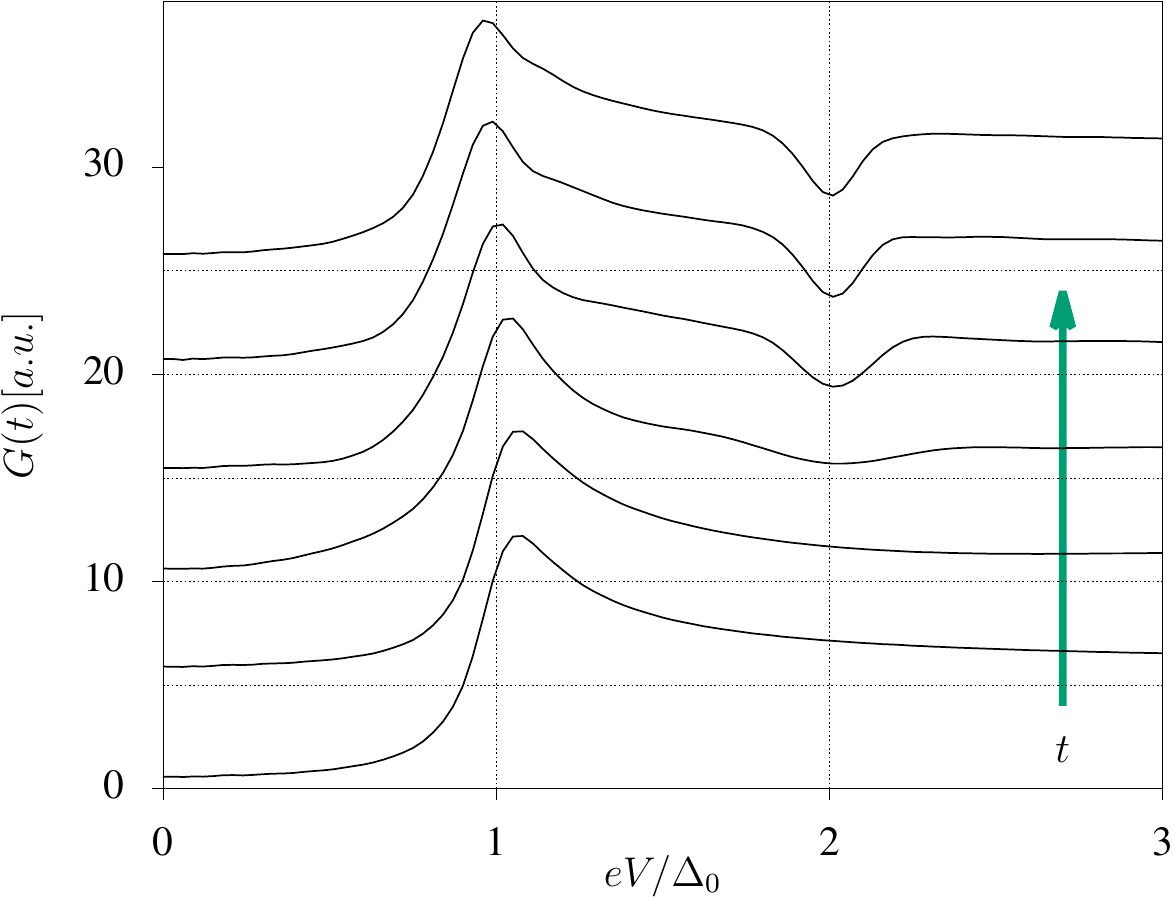}
\caption{Differential conductance for $\alpha=0.1$ at several values of times inside the first period $\tau_R$ of Rabi-Higgs mode, from bottom to top, $t=0$ (equilibrium), $t=0.1\tau_R$, $t=0.25\tau_R$,  $t=0.5\tau_R$,  $t=0.75\tau_R$ and $t=\tau_R$. We use $\gamma=0.2\Delta_{0}$ and the zero for each curve (horizontal dashed line) was displaced by factor 5 for clarity.} 
\label{fig:fig5}
\end{figure}

Turning to the driven case, for small $\gamma$ values the out-of-equilibrium differential conductance is computationally very demanding since the whole dynamics has to be integrated in Eq.~(\ref{eq:conductance}), differently from the photoemission case where the integrals are cutoff by the probe pulse shape. To make the computations feasible we use $\gamma=0.2\Delta_0$ and $\alpha=0.1$. The nominal Rabi-Higgs period corresponds to  $\tau_R\Delta_0\approx27$ but because of the large damping  a steady state is reached before a full Rabi cycle is completed. Still as shown in Fig.~\ref{fig:fig5}, the time-dependent differential conductance clearly shows a non trivial transient dynamics. Indeed, a clear depression in the time-resolved conductance around $eV=2\Delta_0$ can be observed as a function of time which can be identified with  the beginning of the Rabi-Higgs oscillation. 

As in the case of photoemission two effects determine the features in the spectrum. The upper and lower Bogoliubov bands hybridize with the first Floquet sideband of the  lower and upper Bogoliubov bands respectively creating a  pseudogap at $\pm\omega_d/2$ (see right column of Fig~\ref{fig:fig41}). Furthermore, the depression of the population at energy $-\omega_d/2$ depresses the probability to extract electrons and the excess of population at energy $\omega_d/2$ prevents injecting electrons due to Pauli blocking. 

Another effect of the drive is that whole shape of the conductance gets modified. Due to the Dynes parameter, at equilibrium the differential conductance does not vanishes sharply for energies lower than $\Delta_0$. Once the time-dependent perturbation is turned on, the superconducting coherence peak decreases according to a decreasing in the average  order parameter.

\section{Summary and Outlooks}
\label{sec:conc}

We have studied the dynamic of a BCS superconductor subject to a periodic drive in the presence of dissipation focusing on interesting transient dynamics. For $\gamma\lesssim\omega_R$ the transient show   Rabi-Higgs oscillations which persist forever when $\gamma=0$. In the opposite limit $\gamma\gtrsim\omega_R$ oscillations are not observed and the response is practically described by linear response at all times for a weak drive and show the phenomena of saturation of population imbalance for large drives. Another interesting non-linearity is second harmonic generation, which we have found it is allowed for $\lambda$-drive. 

In all regimes the behavior resembles an isolated  driven two level system. However one should keep in mind that the interactions are fundamental in synchronizing a finite fraction of the quasiparticles and make the mode collective. There are close analogies with phenomena in quantum optics where  Rabi oscillations and saturation of population are common phenomena. This analogy paves the way to explore quantum optics protocols to control and manipulate the superconducting state of materials as already proposed in Ref.~\cite{Collado2018}.

The experimental detection of these highly nonlinear behaviors in driven superconductors could be a major step towards the quantum control and manipulation of quantum phases. We have proposed two techniques to detect  Rabi-Higgs oscillations  taking into account dissipation. A cyclic imbalance of quasiparticle population at specific  locations in energy and momentum has been shown to appear in  tr-ARPES. Due to technical reasons only the rise of the population imbalance could be explicitly computed for tunneling experiments but we expect that, in close analogy with photoemission,  
the anomalies oscillate with the Rabi frequency for small dissipation. Even detection of the steady state non-linearities would be a quite interesting experimental achievement. 

The present formalism can be easily extended to take into account more interesting relaxation mechanisms via more sophisticated self-energies beyond the wide-band approximation used here for the bath. An interesting direction is to analyze how robust the Rabi-Higgs mode is in the presence of dephasing, decoherence and relaxation sources from a more microscopic point of view by considering residual Coulomb and electron-phonon interactions where heating effects could be relevant.

\begin{acknowledgments}
J.L. is in debt for enlightening discussions with L.  Benfatto, C. Castellani, G. Seibold, B. Leridon, N. Bergeal and J\'er\^ome Lesueur. We acknowledge financial support from Italian MAECI and Argentinian MINCYT through bilateral project AR17MO7 and from ANPCyT (grants PICTs 2013-1045 and 2016-0791), from CONICET (grant PIP 11220150100506) and from SeCyT-UNCuyo (grant 06/C603). J.L. acknowledges financial support from  Italian MAECI thought collaborative project SUPERTOP-PGR04879, from Italian MIUR though 
Project No. PRIN 2017Z8TS5B, and from Regione Lazio (L. R. 13/08) under project SIMAP.
\end{acknowledgments}


\bibliographystyle{apsrev4-1_title}

\bibliography{library}

\begin{thebibliography}{52}%
\makeatletter
\providecommand \@ifxundefined [1]{%
 \@ifx{#1\undefined}
}%
\providecommand \@ifnum [1]{%
 \ifnum #1\expandafter \@firstoftwo
 \else \expandafter \@secondoftwo
 \fi
}%
\providecommand \@ifx [1]{%
 \ifx #1\expandafter \@firstoftwo
 \else \expandafter \@secondoftwo
 \fi
}%
\providecommand \natexlab [1]{#1}%
\providecommand \enquote  [1]{``#1''}%
\providecommand \bibnamefont  [1]{#1}%
\providecommand \bibfnamefont [1]{#1}%
\providecommand \citenamefont [1]{#1}%
\providecommand \href@noop [0]{\@secondoftwo}%
\providecommand \href [0]{\begingroup \@sanitize@url \@href}%
\providecommand \@href[1]{\@@startlink{#1}\@@href}%
\providecommand \@@href[1]{\endgroup#1\@@endlink}%
\providecommand \@sanitize@url [0]{\catcode `\\12\catcode `\$12\catcode
  `\&12\catcode `\#12\catcode `\^12\catcode `\_12\catcode `\%12\relax}%
\providecommand \@@startlink[1]{}%
\providecommand \@@endlink[0]{}%
\providecommand \url  [0]{\begingroup\@sanitize@url \@url }%
\providecommand \@url [1]{\endgroup\@href {#1}{\urlprefix }}%
\providecommand \urlprefix  [0]{URL }%
\providecommand \Eprint [0]{\href }%
\providecommand \doibase [0]{http://dx.doi.org/}%
\providecommand \selectlanguage [0]{\@gobble}%
\providecommand \bibinfo  [0]{\@secondoftwo}%
\providecommand \bibfield  [0]{\@secondoftwo}%
\providecommand \translation [1]{[#1]}%
\providecommand \BibitemOpen [0]{}%
\providecommand \bibitemStop [0]{}%
\providecommand \bibitemNoStop [0]{.\EOS\space}%
\providecommand \EOS [0]{\spacefactor3000\relax}%
\providecommand \BibitemShut  [1]{\csname bibitem#1\endcsname}%
\let\auto@bib@innerbib\@empty
\bibitem [{\citenamefont {Fausti}\ \emph {et~al.}(2011)\citenamefont {Fausti},
  \citenamefont {Tobey}, \citenamefont {Dean}, \citenamefont {Kaiser},
  \citenamefont {Dienst}, \citenamefont {Hoffmann}, \citenamefont {Pyon},
  \citenamefont {Takayama}, \citenamefont {Takagi},\ and\ \citenamefont
  {Cavalleri}}]{Fausti2011d}%
  \BibitemOpen
  \bibfield  {author} {\bibinfo {author} {\bibfnamefont {D.}~\bibnamefont
  {Fausti}}, \bibinfo {author} {\bibfnamefont {R.~I.}\ \bibnamefont {Tobey}},
  \bibinfo {author} {\bibfnamefont {N.}~\bibnamefont {Dean}}, \bibinfo {author}
  {\bibfnamefont {S.}~\bibnamefont {Kaiser}}, \bibinfo {author} {\bibfnamefont
  {A.}~\bibnamefont {Dienst}}, \bibinfo {author} {\bibfnamefont {M.~C.}\
  \bibnamefont {Hoffmann}}, \bibinfo {author} {\bibfnamefont {S.}~\bibnamefont
  {Pyon}}, \bibinfo {author} {\bibfnamefont {T.}~\bibnamefont {Takayama}},
  \bibinfo {author} {\bibfnamefont {H.}~\bibnamefont {Takagi}}, \ and\ \bibinfo
  {author} {\bibfnamefont {A.}~\bibnamefont {Cavalleri}},\ }\bibfield  {title}
  {\enquote {\bibinfo {title} {{Light-induced superconductivity in a
  stripe-ordered cuprate.}}}\ }\href {\doibase 10.1126/science.1197294}
  {\bibfield  {journal} {\bibinfo  {journal} {Science}\ }\textbf {\bibinfo
  {volume} {331}},\ \bibinfo {pages} {189} (\bibinfo {year}
  {2011})}\BibitemShut {NoStop}%
\bibitem [{\citenamefont {Mansart}\ \emph {et~al.}(2013)\citenamefont
  {Mansart}, \citenamefont {Lorenzana}, \citenamefont {Mann}, \citenamefont
  {Odeh}, \citenamefont {Scarongella}, \citenamefont {Chergui},\ and\
  \citenamefont {Carbone}}]{Mansart2013}%
  \BibitemOpen
  \bibfield  {author} {\bibinfo {author} {\bibfnamefont {B.}~\bibnamefont
  {Mansart}}, \bibinfo {author} {\bibfnamefont {J.}~\bibnamefont {Lorenzana}},
  \bibinfo {author} {\bibfnamefont {a.}~\bibnamefont {Mann}}, \bibinfo {author}
  {\bibfnamefont {a.}~\bibnamefont {Odeh}}, \bibinfo {author} {\bibfnamefont
  {M.}~\bibnamefont {Scarongella}}, \bibinfo {author} {\bibfnamefont
  {M.}~\bibnamefont {Chergui}}, \ and\ \bibinfo {author} {\bibfnamefont
  {F.}~\bibnamefont {Carbone}},\ }\bibfield  {title} {\enquote {\bibinfo
  {title} {{Coupling of a high-energy excitation to superconducting
  quasiparticles in a cuprate from coherent charge fluctuation
  spectroscopy}},}\ }\href {\doibase 10.1073/pnas.1218742110} {\bibfield
  {journal} {\bibinfo  {journal} {Proc. Natl. Acad. Sci.}\ }\textbf {\bibinfo
  {volume} {110}},\ \bibinfo {pages} {4539} (\bibinfo {year}
  {2013})}\BibitemShut {NoStop}%
\bibitem [{\citenamefont {Matsunaga}\ \emph {et~al.}(2013)\citenamefont
  {Matsunaga}, \citenamefont {Hamada}, \citenamefont {Makise}, \citenamefont
  {Uzawa}, \citenamefont {Terai}, \citenamefont {Wang},\ and\ \citenamefont
  {Shimano}}]{Matsunaga2013}%
  \BibitemOpen
  \bibfield  {author} {\bibinfo {author} {\bibfnamefont {R.}~\bibnamefont
  {Matsunaga}}, \bibinfo {author} {\bibfnamefont {Y.~I.}\ \bibnamefont
  {Hamada}}, \bibinfo {author} {\bibfnamefont {K.}~\bibnamefont {Makise}},
  \bibinfo {author} {\bibfnamefont {Y.}~\bibnamefont {Uzawa}}, \bibinfo
  {author} {\bibfnamefont {H.}~\bibnamefont {Terai}}, \bibinfo {author}
  {\bibfnamefont {Z.}~\bibnamefont {Wang}}, \ and\ \bibinfo {author}
  {\bibfnamefont {R.}~\bibnamefont {Shimano}},\ }\bibfield  {title} {\enquote
  {\bibinfo {title} {{Higgs amplitude mode in the {BCS} superconductors Nb1-xTi
  xN induced by terahertz pulse excitation}},}\ }\href {\doibase
  10.1103/PhysRevLett.111.057002} {\bibfield  {journal} {\bibinfo  {journal}
  {Phys. Rev. Lett.}\ }\textbf {\bibinfo {volume} {111}},\ \bibinfo {pages}
  {057002} (\bibinfo {year} {2013})}\BibitemShut {NoStop}%
\bibitem [{\citenamefont {Matsunaga}\ \emph {et~al.}(2014)\citenamefont
  {Matsunaga}, \citenamefont {Tsuji}, \citenamefont {Fujita}, \citenamefont
  {Sugioka}, \citenamefont {Makise}, \citenamefont {Uzawa}, \citenamefont
  {Terai}, \citenamefont {Wang}, \citenamefont {Aoki},\ and\ \citenamefont
  {Shimano}}]{Matsunaga2014}%
  \BibitemOpen
  \bibfield  {author} {\bibinfo {author} {\bibfnamefont {R.}~\bibnamefont
  {Matsunaga}}, \bibinfo {author} {\bibfnamefont {N.}~\bibnamefont {Tsuji}},
  \bibinfo {author} {\bibfnamefont {H.}~\bibnamefont {Fujita}}, \bibinfo
  {author} {\bibfnamefont {A.}~\bibnamefont {Sugioka}}, \bibinfo {author}
  {\bibfnamefont {K.}~\bibnamefont {Makise}}, \bibinfo {author} {\bibfnamefont
  {Y.}~\bibnamefont {Uzawa}}, \bibinfo {author} {\bibfnamefont
  {H.}~\bibnamefont {Terai}}, \bibinfo {author} {\bibfnamefont
  {Z.}~\bibnamefont {Wang}}, \bibinfo {author} {\bibfnamefont {H.}~\bibnamefont
  {Aoki}}, \ and\ \bibinfo {author} {\bibfnamefont {R.}~\bibnamefont
  {Shimano}},\ }\bibfield  {title} {\enquote {\bibinfo {title} {{Light-induced
  collective pseudospin precession resonating with Higgs mode in a
  superconductor}},}\ }\href {\doibase 10.1126/science.1254697} {\bibfield
  {journal} {\bibinfo  {journal} {Science}\ }\textbf {\bibinfo {volume}
  {345}},\ \bibinfo {pages} {1145} (\bibinfo {year} {2014})}\BibitemShut
  {NoStop}%
\bibitem [{\citenamefont {Mankowsky}\ \emph {et~al.}(2014)\citenamefont
  {Mankowsky}, \citenamefont {Subedi}, \citenamefont {F{\"{o}}rst},
  \citenamefont {Mariager}, \citenamefont {Chollet}, \citenamefont {Lemke},
  \citenamefont {Robinson}, \citenamefont {Glownia}, \citenamefont {Minitti},
  \citenamefont {Frano}, \citenamefont {Fechner}, \citenamefont {Spaldin},
  \citenamefont {Loew}, \citenamefont {Keimer}, \citenamefont {Georges},\ and\
  \citenamefont {Cavalleri}}]{Mankowsky2014}%
  \BibitemOpen
  \bibfield  {author} {\bibinfo {author} {\bibfnamefont {R.}~\bibnamefont
  {Mankowsky}}, \bibinfo {author} {\bibfnamefont {a.}~\bibnamefont {Subedi}},
  \bibinfo {author} {\bibfnamefont {M.}~\bibnamefont {F{\"{o}}rst}}, \bibinfo
  {author} {\bibfnamefont {S.~O.}\ \bibnamefont {Mariager}}, \bibinfo {author}
  {\bibfnamefont {M.}~\bibnamefont {Chollet}}, \bibinfo {author} {\bibfnamefont
  {H.~T.}\ \bibnamefont {Lemke}}, \bibinfo {author} {\bibfnamefont {J.~S.}\
  \bibnamefont {Robinson}}, \bibinfo {author} {\bibfnamefont {J.~M.}\
  \bibnamefont {Glownia}}, \bibinfo {author} {\bibfnamefont {M.~P.}\
  \bibnamefont {Minitti}}, \bibinfo {author} {\bibfnamefont {A.}~\bibnamefont
  {Frano}}, \bibinfo {author} {\bibfnamefont {M.}~\bibnamefont {Fechner}},
  \bibinfo {author} {\bibfnamefont {N.~A.}\ \bibnamefont {Spaldin}}, \bibinfo
  {author} {\bibfnamefont {T.}~\bibnamefont {Loew}}, \bibinfo {author}
  {\bibfnamefont {B.}~\bibnamefont {Keimer}}, \bibinfo {author} {\bibfnamefont
  {A.}~\bibnamefont {Georges}}, \ and\ \bibinfo {author} {\bibfnamefont
  {A.}~\bibnamefont {Cavalleri}},\ }\bibfield  {title} {\enquote {\bibinfo
  {title} {{Nonlinear lattice dynamics as a basis for enhanced
  superconductivity in YBa2Cu3O6.5}},}\ }\href {\doibase 10.1038/nature13875}
  {\bibfield  {journal} {\bibinfo  {journal} {Nature}\ }\textbf {\bibinfo
  {volume} {516}},\ \bibinfo {pages} {71} (\bibinfo {year} {2014})}\BibitemShut
  {NoStop}%
\bibitem [{\citenamefont {Nicoletti}\ \emph {et~al.}(2014)\citenamefont
  {Nicoletti}, \citenamefont {Casandruc}, \citenamefont {Laplace},
  \citenamefont {Khanna}, \citenamefont {Hunt}, \citenamefont {Kaiser},
  \citenamefont {Dhesi}, \citenamefont {Gu}, \citenamefont {Hill},\ and\
  \citenamefont {Cavalleri}}]{Nicoletti2014}%
  \BibitemOpen
  \bibfield  {author} {\bibinfo {author} {\bibfnamefont {D.}~\bibnamefont
  {Nicoletti}}, \bibinfo {author} {\bibfnamefont {E.}~\bibnamefont
  {Casandruc}}, \bibinfo {author} {\bibfnamefont {Y.}~\bibnamefont {Laplace}},
  \bibinfo {author} {\bibfnamefont {V.}~\bibnamefont {Khanna}}, \bibinfo
  {author} {\bibfnamefont {C.~R.}\ \bibnamefont {Hunt}}, \bibinfo {author}
  {\bibfnamefont {S.}~\bibnamefont {Kaiser}}, \bibinfo {author} {\bibfnamefont
  {S.~S.}\ \bibnamefont {Dhesi}}, \bibinfo {author} {\bibfnamefont {G.~D.}\
  \bibnamefont {Gu}}, \bibinfo {author} {\bibfnamefont {J.~P.}\ \bibnamefont
  {Hill}}, \ and\ \bibinfo {author} {\bibfnamefont {A.}~\bibnamefont
  {Cavalleri}},\ }\bibfield  {title} {\enquote {\bibinfo {title} {{Optically
  induced superconductivity in striped La2-xBaxCuO4 by polarization-selective
  excitation in the near infrared}},}\ }\href {\doibase
  10.1103/PhysRevB.90.100503} {\bibfield  {journal} {\bibinfo  {journal} {Phys.
  Rev. B}\ }\textbf {\bibinfo {volume} {90}},\ \bibinfo {pages} {1} (\bibinfo
  {year} {2014})}\BibitemShut {NoStop}%
\bibitem [{\citenamefont {Kaiser}\ \emph {et~al.}(2014)\citenamefont {Kaiser},
  \citenamefont {Hunt}, \citenamefont {Nicoletti}, \citenamefont {Hu},
  \citenamefont {Gierz}, \citenamefont {Liu}, \citenamefont {{Le Tacon}},
  \citenamefont {Loew}, \citenamefont {Haug}, \citenamefont {Keimer},\ and\
  \citenamefont {Cavalleri}}]{Kaiser2014}%
  \BibitemOpen
  \bibfield  {author} {\bibinfo {author} {\bibfnamefont {S.}~\bibnamefont
  {Kaiser}}, \bibinfo {author} {\bibfnamefont {C.~R.}\ \bibnamefont {Hunt}},
  \bibinfo {author} {\bibfnamefont {D.}~\bibnamefont {Nicoletti}}, \bibinfo
  {author} {\bibfnamefont {W.}~\bibnamefont {Hu}}, \bibinfo {author}
  {\bibfnamefont {I.}~\bibnamefont {Gierz}}, \bibinfo {author} {\bibfnamefont
  {H.~Y.}\ \bibnamefont {Liu}}, \bibinfo {author} {\bibfnamefont
  {M.}~\bibnamefont {{Le Tacon}}}, \bibinfo {author} {\bibfnamefont
  {T.}~\bibnamefont {Loew}}, \bibinfo {author} {\bibfnamefont {D.}~\bibnamefont
  {Haug}}, \bibinfo {author} {\bibfnamefont {B.}~\bibnamefont {Keimer}}, \ and\
  \bibinfo {author} {\bibfnamefont {A.}~\bibnamefont {Cavalleri}},\ }\bibfield
  {title} {\enquote {\bibinfo {title} {{Optically induced coherent transport
  far above Tc in underdoped YBa2Cu3 O6+??}}}\ }\href {\doibase
  10.1103/PhysRevB.89.184516} {\bibfield  {journal} {\bibinfo  {journal} {Phys.
  Rev. B}\ }\textbf {\bibinfo {volume} {89}},\ \bibinfo {pages} {1} (\bibinfo
  {year} {2014})},\ \Eprint {http://arxiv.org/abs/1205.4661} {arXiv:1205.4661}
  \BibitemShut {NoStop}%
\bibitem [{\citenamefont {Mitrano}\ \emph {et~al.}(2016)\citenamefont
  {Mitrano}, \citenamefont {Cantaluppi}, \citenamefont {Nicoletti},
  \citenamefont {Kaiser}, \citenamefont {Perucchi}, \citenamefont {Lupi},
  \citenamefont {{Di Pietro}}, \citenamefont {Pontiroli}, \citenamefont
  {Ricc{\`{o}}}, \citenamefont {Clark}, \citenamefont {Jaksch},\ and\
  \citenamefont {Cavalleri}}]{Mitrano2016}%
  \BibitemOpen
  \bibfield  {author} {\bibinfo {author} {\bibfnamefont {M.}~\bibnamefont
  {Mitrano}}, \bibinfo {author} {\bibfnamefont {A.}~\bibnamefont {Cantaluppi}},
  \bibinfo {author} {\bibfnamefont {D.}~\bibnamefont {Nicoletti}}, \bibinfo
  {author} {\bibfnamefont {S.}~\bibnamefont {Kaiser}}, \bibinfo {author}
  {\bibfnamefont {A.}~\bibnamefont {Perucchi}}, \bibinfo {author}
  {\bibfnamefont {S.}~\bibnamefont {Lupi}}, \bibinfo {author} {\bibfnamefont
  {P.}~\bibnamefont {{Di Pietro}}}, \bibinfo {author} {\bibfnamefont
  {D.}~\bibnamefont {Pontiroli}}, \bibinfo {author} {\bibfnamefont
  {M.}~\bibnamefont {Ricc{\`{o}}}}, \bibinfo {author} {\bibfnamefont {S.~R.}\
  \bibnamefont {Clark}}, \bibinfo {author} {\bibfnamefont {D.}~\bibnamefont
  {Jaksch}}, \ and\ \bibinfo {author} {\bibfnamefont {A.}~\bibnamefont
  {Cavalleri}},\ }\bibfield  {title} {\enquote {\bibinfo {title} {{Possible
  light-induced superconductivity in K3 C60 at high temperature}},}\ }\href
  {\doibase 10.1038/nature16522} {\bibfield  {journal} {\bibinfo  {journal}
  {Nature}\ }\textbf {\bibinfo {volume} {530}},\ \bibinfo {pages} {461}
  (\bibinfo {year} {2016})}\BibitemShut {NoStop}%
\bibitem [{\citenamefont {Rajasekaran}\ \emph {et~al.}(2018)\citenamefont
  {Rajasekaran}, \citenamefont {Okamoto}, \citenamefont {Mathey}, \citenamefont
  {Fechner}, \citenamefont {Thampy}, \citenamefont {Gu},\ and\ \citenamefont
  {Cavalleri}}]{Rajasekaran2018}%
  \BibitemOpen
  \bibfield  {author} {\bibinfo {author} {\bibfnamefont {S.}~\bibnamefont
  {Rajasekaran}}, \bibinfo {author} {\bibfnamefont {J.}~\bibnamefont
  {Okamoto}}, \bibinfo {author} {\bibfnamefont {L.}~\bibnamefont {Mathey}},
  \bibinfo {author} {\bibfnamefont {M.}~\bibnamefont {Fechner}}, \bibinfo
  {author} {\bibfnamefont {V.}~\bibnamefont {Thampy}}, \bibinfo {author}
  {\bibfnamefont {G.~D.}\ \bibnamefont {Gu}}, \ and\ \bibinfo {author}
  {\bibfnamefont {A.}~\bibnamefont {Cavalleri}},\ }\bibfield  {title} {\enquote
  {\bibinfo {title} {{Probing optically silent superfluid stripes in
  cuprates}},}\ }\href {\doibase 10.1126/science.aan3438} {\bibfield  {journal}
  {\bibinfo  {journal} {Science (80-. ).}\ }\textbf {\bibinfo {volume} {359}},\
  \bibinfo {pages} {575} (\bibinfo {year} {2018})}\BibitemShut {NoStop}%
\bibitem [{\citenamefont {St{\"{o}}ferle}\ \emph {et~al.}(2004)\citenamefont
  {St{\"{o}}ferle}, \citenamefont {Moritz}, \citenamefont {Schori},
  \citenamefont {K{\"{o}}hl},\ and\ \citenamefont {Esslinger}}]{Stoferle2004}%
  \BibitemOpen
  \bibfield  {author} {\bibinfo {author} {\bibfnamefont {T.}~\bibnamefont
  {St{\"{o}}ferle}}, \bibinfo {author} {\bibfnamefont {H.}~\bibnamefont
  {Moritz}}, \bibinfo {author} {\bibfnamefont {C.}~\bibnamefont {Schori}},
  \bibinfo {author} {\bibfnamefont {M.}~\bibnamefont {K{\"{o}}hl}}, \ and\
  \bibinfo {author} {\bibfnamefont {T.}~\bibnamefont {Esslinger}},\ }\bibfield
  {title} {\enquote {\bibinfo {title} {{Transition from a Strongly Interacting
  1D Superfluid to a Mott Insulator}},}\ }\href {\doibase
  10.1103/PhysRevLett.92.130403} {\bibfield  {journal} {\bibinfo  {journal}
  {Phys. Rev. Lett.}\ }\textbf {\bibinfo {volume} {92}},\ \bibinfo {pages}
  {130403} (\bibinfo {year} {2004})}\BibitemShut {NoStop}%
\bibitem [{\citenamefont {Haller}\ \emph {et~al.}(2010)\citenamefont {Haller},
  \citenamefont {Hart}, \citenamefont {Mark}, \citenamefont {Danzl},
  \citenamefont {Reichs{\"{o}}llner}, \citenamefont {Gustavsson}, \citenamefont
  {Dalmonte}, \citenamefont {Pupillo},\ and\ \citenamefont
  {N{\"{a}}gerl}}]{Haller2010}%
  \BibitemOpen
  \bibfield  {author} {\bibinfo {author} {\bibfnamefont {E.}~\bibnamefont
  {Haller}}, \bibinfo {author} {\bibfnamefont {R.}~\bibnamefont {Hart}},
  \bibinfo {author} {\bibfnamefont {M.~J.}\ \bibnamefont {Mark}}, \bibinfo
  {author} {\bibfnamefont {J.~G.}\ \bibnamefont {Danzl}}, \bibinfo {author}
  {\bibfnamefont {L.}~\bibnamefont {Reichs{\"{o}}llner}}, \bibinfo {author}
  {\bibfnamefont {M.}~\bibnamefont {Gustavsson}}, \bibinfo {author}
  {\bibfnamefont {M.}~\bibnamefont {Dalmonte}}, \bibinfo {author}
  {\bibfnamefont {G.}~\bibnamefont {Pupillo}}, \ and\ \bibinfo {author}
  {\bibfnamefont {H.~C.}\ \bibnamefont {N{\"{a}}gerl}},\ }\bibfield  {title}
  {\enquote {\bibinfo {title} {{Pinning quantum phase transition for a
  Luttinger liquid of strongly interacting bosons}},}\ }\href {\doibase
  10.1038/nature09259} {\bibfield  {journal} {\bibinfo  {journal} {Nature}\
  }\textbf {\bibinfo {volume} {466}},\ \bibinfo {pages} {597} (\bibinfo {year}
  {2010})}\BibitemShut {NoStop}%
\bibitem [{\citenamefont {Endres}\ \emph {et~al.}(2012)\citenamefont {Endres},
  \citenamefont {Fukuhara}, \citenamefont {Pekker}, \citenamefont {Cheneau},
  \citenamefont {Schau{\ss}}, \citenamefont {Gross}, \citenamefont {Demler},
  \citenamefont {Kuhr},\ and\ \citenamefont {Bloch}}]{Endres2012}%
  \BibitemOpen
  \bibfield  {author} {\bibinfo {author} {\bibfnamefont {M.}~\bibnamefont
  {Endres}}, \bibinfo {author} {\bibfnamefont {T.}~\bibnamefont {Fukuhara}},
  \bibinfo {author} {\bibfnamefont {D.}~\bibnamefont {Pekker}}, \bibinfo
  {author} {\bibfnamefont {M.}~\bibnamefont {Cheneau}}, \bibinfo {author}
  {\bibfnamefont {P.}~\bibnamefont {Schau{\ss}}}, \bibinfo {author}
  {\bibfnamefont {C.}~\bibnamefont {Gross}}, \bibinfo {author} {\bibfnamefont
  {E.}~\bibnamefont {Demler}}, \bibinfo {author} {\bibfnamefont
  {S.}~\bibnamefont {Kuhr}}, \ and\ \bibinfo {author} {\bibfnamefont
  {I.}~\bibnamefont {Bloch}},\ }\bibfield  {title} {\enquote {\bibinfo {title}
  {{The 'Higgs' amplitude mode at the two-dimensional superfluid/Mott insulator
  transition}},}\ }\href {\doibase 10.1038/nature11255} {\bibfield  {journal}
  {\bibinfo  {journal} {Nature}\ }\textbf {\bibinfo {volume} {487}},\ \bibinfo
  {pages} {454} (\bibinfo {year} {2012})}\BibitemShut {NoStop}%
\bibitem [{\citenamefont {Chin}\ \emph {et~al.}(2010)\citenamefont {Chin},
  \citenamefont {Grimm}, \citenamefont {Julienne},\ and\ \citenamefont
  {Tiesinga}}]{Chin2010}%
  \BibitemOpen
  \bibfield  {author} {\bibinfo {author} {\bibfnamefont {C.}~\bibnamefont
  {Chin}}, \bibinfo {author} {\bibfnamefont {R.}~\bibnamefont {Grimm}},
  \bibinfo {author} {\bibfnamefont {P.}~\bibnamefont {Julienne}}, \ and\
  \bibinfo {author} {\bibfnamefont {E.}~\bibnamefont {Tiesinga}},\ }\bibfield
  {title} {\enquote {\bibinfo {title} {{Feshbach resonances in ultracold
  gases}},}\ }\href {\doibase 10.1103/RevModPhys.82.1225} {\bibfield  {journal}
  {\bibinfo  {journal} {Rev. Mod. Phys.}\ }\textbf {\bibinfo {volume} {82}},\
  \bibinfo {pages} {1225} (\bibinfo {year} {2010})}\BibitemShut {NoStop}%
\bibitem [{\citenamefont {Behrle}\ \emph {et~al.}(2018)\citenamefont {Behrle},
  \citenamefont {Harrison}, \citenamefont {Kombe}, \citenamefont {Gao},
  \citenamefont {Link}, \citenamefont {Bernier}, \citenamefont {Kollath},\ and\
  \citenamefont {K{\"{o}}hl}}]{Behrle2018}%
  \BibitemOpen
  \bibfield  {author} {\bibinfo {author} {\bibfnamefont {A.}~\bibnamefont
  {Behrle}}, \bibinfo {author} {\bibfnamefont {T.}~\bibnamefont {Harrison}},
  \bibinfo {author} {\bibfnamefont {J.}~\bibnamefont {Kombe}}, \bibinfo
  {author} {\bibfnamefont {K.}~\bibnamefont {Gao}}, \bibinfo {author}
  {\bibfnamefont {M.}~\bibnamefont {Link}}, \bibinfo {author} {\bibfnamefont
  {J.-S.}\ \bibnamefont {Bernier}}, \bibinfo {author} {\bibfnamefont
  {C.}~\bibnamefont {Kollath}}, \ and\ \bibinfo {author} {\bibfnamefont
  {M.}~\bibnamefont {K{\"{o}}hl}},\ }\bibfield  {title} {\enquote {\bibinfo
  {title} {{Higgs mode in a strongly interacting fermionic superfluid}},}\
  }\href {\doibase 10.1038/s41567-018-0128-6} {\bibfield  {journal} {\bibinfo
  {journal} {Nat. Phys.}\ }\textbf {\bibinfo {volume} {14}},\ \bibinfo {pages}
  {781} (\bibinfo {year} {2018})}\BibitemShut {NoStop}%
\bibitem [{\citenamefont {Clark}\ \emph {et~al.}(2015)\citenamefont {Clark},
  \citenamefont {Ha}, \citenamefont {Xu},\ and\ \citenamefont
  {Chin}}]{Clark2015}%
  \BibitemOpen
  \bibfield  {author} {\bibinfo {author} {\bibfnamefont {L.~W.}\ \bibnamefont
  {Clark}}, \bibinfo {author} {\bibfnamefont {L.-C.}\ \bibnamefont {Ha}},
  \bibinfo {author} {\bibfnamefont {C.-Y.}\ \bibnamefont {Xu}}, \ and\ \bibinfo
  {author} {\bibfnamefont {C.}~\bibnamefont {Chin}},\ }\bibfield  {title}
  {\enquote {\bibinfo {title} {{Quantum Dynamics with Spatiotemporal Control of
  Interactions in a Stable Bose-Einstein Condensate}},}\ }\href {\doibase
  10.1103/PhysRevLett.115.155301} {\bibfield  {journal} {\bibinfo  {journal}
  {Phys. Rev. Lett.}\ }\textbf {\bibinfo {volume} {115}},\ \bibinfo {pages}
  {155301} (\bibinfo {year} {2015})}\BibitemShut {NoStop}%
\bibitem [{\citenamefont {Barankov}\ and\ \citenamefont
  {Levitov}(2006)}]{Barankov2006a}%
  \BibitemOpen
  \bibfield  {author} {\bibinfo {author} {\bibfnamefont {R.~A.}\ \bibnamefont
  {Barankov}}\ and\ \bibinfo {author} {\bibfnamefont {L.~S.}\ \bibnamefont
  {Levitov}},\ }\bibfield  {title} {\enquote {\bibinfo {title}
  {{Synchronization in the BCS pairing dynamics as a critical phenomenon}},}\
  }\href {\doibase 10.1103/PhysRevLett.96.230403} {\bibfield  {journal}
  {\bibinfo  {journal} {Phys. Rev. Lett.}\ }\textbf {\bibinfo {volume} {96}},\
  \bibinfo {pages} {1} (\bibinfo {year} {2006})}\BibitemShut {NoStop}%
\bibitem [{\citenamefont {Eckstein}\ \emph {et~al.}(2009)\citenamefont
  {Eckstein}, \citenamefont {Kollar},\ and\ \citenamefont
  {Werner}}]{Eckstein2009}%
  \BibitemOpen
  \bibfield  {author} {\bibinfo {author} {\bibfnamefont {M.}~\bibnamefont
  {Eckstein}}, \bibinfo {author} {\bibfnamefont {M.}~\bibnamefont {Kollar}}, \
  and\ \bibinfo {author} {\bibfnamefont {P.}~\bibnamefont {Werner}},\
  }\bibfield  {title} {\enquote {\bibinfo {title} {{Thermalization after an
  Interaction Quench in the Hubbard Model}},}\ }\href {\doibase
  10.1103/PhysRevLett.103.056403} {\bibfield  {journal} {\bibinfo  {journal}
  {Phys. Rev. Lett.}\ }\textbf {\bibinfo {volume} {103}},\ \bibinfo {pages}
  {056403} (\bibinfo {year} {2009})}\BibitemShut {NoStop}%
\bibitem [{\citenamefont {Collado}\ \emph {et~al.}(2018)\citenamefont
  {Collado}, \citenamefont {Lorenzana}, \citenamefont {Usaj},\ and\
  \citenamefont {Balseiro}}]{Collado2018}%
  \BibitemOpen
  \bibfield  {author} {\bibinfo {author} {\bibfnamefont {H.~P.~O.}\
  \bibnamefont {Collado}}, \bibinfo {author} {\bibfnamefont {J.}~\bibnamefont
  {Lorenzana}}, \bibinfo {author} {\bibfnamefont {G.}~\bibnamefont {Usaj}}, \
  and\ \bibinfo {author} {\bibfnamefont {C.~A.}\ \bibnamefont {Balseiro}},\
  }\bibfield  {title} {\enquote {\bibinfo {title} {{Population inversion and
  dynamical phase transitions in a driven superconductor}},}\ }\href {\doibase
  10.1103/PhysRevB.98.214519} {\bibfield  {journal} {\bibinfo  {journal} {Phys.
  Rev. B}\ }\textbf {\bibinfo {volume} {98}},\ \bibinfo {pages} {214519}
  (\bibinfo {year} {2018})}\BibitemShut {NoStop}%
\bibitem [{\citenamefont {Volkov}\ and\ \citenamefont
  {Kogan}(1974)}]{Volkov1974}%
  \BibitemOpen
  \bibfield  {author} {\bibinfo {author} {\bibfnamefont {A.}~\bibnamefont
  {Volkov}}\ and\ \bibinfo {author} {\bibfnamefont {S.}~\bibnamefont {Kogan}},\
  }\bibfield  {title} {\enquote {\bibinfo {title} {{Collisionless relaxation of
  the energy gap in superconductors}},}\ }\href@noop {} {\bibfield  {journal}
  {\bibinfo  {journal} {Sov. J. Exp. Theor. Phys.}\ }\textbf {\bibinfo {volume}
  {38}},\ \bibinfo {pages} {1018} (\bibinfo {year} {1974})}\BibitemShut
  {NoStop}%
\bibitem [{\citenamefont {Barankov}\ \emph {et~al.}(2004)\citenamefont
  {Barankov}, \citenamefont {Levitov},\ and\ \citenamefont
  {Spivak}}]{Barankov2004}%
  \BibitemOpen
  \bibfield  {author} {\bibinfo {author} {\bibfnamefont {R.~A.}\ \bibnamefont
  {Barankov}}, \bibinfo {author} {\bibfnamefont {L.~S.}\ \bibnamefont
  {Levitov}}, \ and\ \bibinfo {author} {\bibfnamefont {B.~Z.}\ \bibnamefont
  {Spivak}},\ }\bibfield  {title} {\enquote {\bibinfo {title} {{Collective Rabi
  oscillations and solitons in a time-dependent BCS pairing problem}},}\ }\href
  {\doibase 10.1103/PhysRevLett.93.160401} {\bibfield  {journal} {\bibinfo
  {journal} {Phys. Rev. Lett.}\ }\textbf {\bibinfo {volume} {93}},\ \bibinfo
  {pages} {160401} (\bibinfo {year} {2004})}\BibitemShut {NoStop}%
\bibitem [{\citenamefont {Lorenzana}\ \emph {et~al.}(2013)\citenamefont
  {Lorenzana}, \citenamefont {Mansart}, \citenamefont {Mann}, \citenamefont
  {Odeh}, \citenamefont {Chergui},\ and\ \citenamefont
  {Carbone}}]{Lorenzana2013}%
  \BibitemOpen
  \bibfield  {author} {\bibinfo {author} {\bibfnamefont {J.}~\bibnamefont
  {Lorenzana}}, \bibinfo {author} {\bibfnamefont {B.}~\bibnamefont {Mansart}},
  \bibinfo {author} {\bibfnamefont {A.}~\bibnamefont {Mann}}, \bibinfo {author}
  {\bibfnamefont {A.}~\bibnamefont {Odeh}}, \bibinfo {author} {\bibfnamefont
  {M.}~\bibnamefont {Chergui}}, \ and\ \bibinfo {author} {\bibfnamefont
  {F.}~\bibnamefont {Carbone}},\ }\bibfield  {title} {\enquote {\bibinfo
  {title} {{Investigating pairing interactions with coherent charge fluctuation
  spectroscopy}},}\ }\href {http://dx.doi.org/10.1140/epjst%2Fe2013-01917-9}
  {\bibfield  {journal} {\bibinfo  {journal} {Eur. Phys. J. Spec. Top.}\
  }\textbf {\bibinfo {volume} {222}},\ \bibinfo {pages} {1223} (\bibinfo {year}
  {2013})}\BibitemShut {NoStop}%
\bibitem [{\citenamefont {Balseiro}\ and\ \citenamefont
  {Falicov}(1980)}]{Balseiro1980}%
  \BibitemOpen
  \bibfield  {author} {\bibinfo {author} {\bibfnamefont {C.~A.}\ \bibnamefont
  {Balseiro}}\ and\ \bibinfo {author} {\bibfnamefont {L.~M.}\ \bibnamefont
  {Falicov}},\ }\bibfield  {title} {\enquote {\bibinfo {title} {{Phonon Raman
  scattering in superconductors}},}\ }\href {\doibase
  10.1103/PhysRevLett.45.662} {\bibfield  {journal} {\bibinfo  {journal} {Phys.
  Rev. Lett.}\ }\textbf {\bibinfo {volume} {45}},\ \bibinfo {pages} {662}
  (\bibinfo {year} {1980})}\BibitemShut {NoStop}%
\bibitem [{\citenamefont {Littlewood}\ and\ \citenamefont
  {Varma}(1982)}]{Littlewood1982}%
  \BibitemOpen
  \bibfield  {author} {\bibinfo {author} {\bibfnamefont {P.~B.}\ \bibnamefont
  {Littlewood}}\ and\ \bibinfo {author} {\bibfnamefont {C.~M.}\ \bibnamefont
  {Varma}},\ }\bibfield  {title} {\enquote {\bibinfo {title} {{Amplitude
  collective modes in superconductors and their coupling to charge-density
  waves}},}\ }\href {\doibase 10.1103/PhysRevB.26.4883} {\bibfield  {journal}
  {\bibinfo  {journal} {Phys. Rev. B}\ }\textbf {\bibinfo {volume} {26}},\
  \bibinfo {pages} {4883} (\bibinfo {year} {1982})}\BibitemShut {NoStop}%
\bibitem [{\citenamefont {Cea}\ \emph {et~al.}(2016)\citenamefont {Cea},
  \citenamefont {Castellani},\ and\ \citenamefont {Benfatto}}]{Cea2015}%
  \BibitemOpen
  \bibfield  {author} {\bibinfo {author} {\bibfnamefont {T.}~\bibnamefont
  {Cea}}, \bibinfo {author} {\bibfnamefont {C.}~\bibnamefont {Castellani}}, \
  and\ \bibinfo {author} {\bibfnamefont {L.}~\bibnamefont {Benfatto}},\
  }\bibfield  {title} {\enquote {\bibinfo {title} {Nonlinear optical effects
  and third-harmonic generation in superconductors: Cooper pairs versus higgs
  mode contribution},}\ }\href {\doibase 10.1103/PhysRevB.93.180507} {\bibfield
   {journal} {\bibinfo  {journal} {Phys. Rev. B}\ }\textbf {\bibinfo {volume}
  {93}},\ \bibinfo {pages} {180507} (\bibinfo {year} {2016})}\BibitemShut
  {NoStop}%
\bibitem [{\citenamefont {Steck}(2019)}]{Steck2019}%
  \BibitemOpen
  \bibfield  {author} {\bibinfo {author} {\bibfnamefont {D.~A.}\ \bibnamefont
  {Steck}},\ }\href@noop {} {\enquote {\bibinfo {title} {{Quantum and Atom
  Optics}},}\ } (\bibinfo {year} {2019})\BibitemShut {NoStop}%
\bibitem [{\citenamefont {Mishchenko}(2009)}]{Mishchenko2009}%
  \BibitemOpen
  \bibfield  {author} {\bibinfo {author} {\bibfnamefont {E.~G.}\ \bibnamefont
  {Mishchenko}},\ }\bibfield  {title} {\enquote {\bibinfo {title} {Dynamic
  conductivity in graphene beyond linear response},}\ }\href {\doibase
  10.1103/PhysRevLett.103.246802} {\bibfield  {journal} {\bibinfo  {journal}
  {Phys. Rev. Lett.}\ }\textbf {\bibinfo {volume} {103}},\ \bibinfo {pages}
  {246802} (\bibinfo {year} {2009})}\BibitemShut {NoStop}%
\bibitem [{\citenamefont {Dynes}\ \emph {et~al.}(1978)\citenamefont {Dynes},
  \citenamefont {Narayanamurti},\ and\ \citenamefont {Garno}}]{Dynes1978}%
  \BibitemOpen
  \bibfield  {author} {\bibinfo {author} {\bibfnamefont {R.~C.}\ \bibnamefont
  {Dynes}}, \bibinfo {author} {\bibfnamefont {V.}~\bibnamefont
  {Narayanamurti}}, \ and\ \bibinfo {author} {\bibfnamefont {J.~P.}\
  \bibnamefont {Garno}},\ }\bibfield  {title} {\enquote {\bibinfo {title}
  {{Direct Measurement of Quasiparticle-Lifetime Broadening in a Strong-Coupled
  Superconductor}},}\ }\href {\doibase 10.1103/PhysRevLett.41.1509} {\bibfield
  {journal} {\bibinfo  {journal} {Phys. Rev. Lett.}\ }\textbf {\bibinfo
  {volume} {41}},\ \bibinfo {pages} {1509} (\bibinfo {year}
  {1978})}\BibitemShut {NoStop}%
\bibitem [{\citenamefont {Collado}\ \emph {et~al.}(2019)\citenamefont
  {Collado}, \citenamefont {Usaj}, \citenamefont {Lorenzana},\ and\
  \citenamefont {Balseiro}}]{Collado2019}%
  \BibitemOpen
  \bibfield  {author} {\bibinfo {author} {\bibfnamefont {H.~P.~O.}\
  \bibnamefont {Collado}}, \bibinfo {author} {\bibfnamefont {G.}~\bibnamefont
  {Usaj}}, \bibinfo {author} {\bibfnamefont {J.}~\bibnamefont {Lorenzana}}, \
  and\ \bibinfo {author} {\bibfnamefont {C.~A.}\ \bibnamefont {Balseiro}},\
  }\bibfield  {title} {\enquote {\bibinfo {title} {{Fate of dynamical phases of
  a BCS superconductor beyond the dissipationless regimen}},}\ }\href {\doibase
  10.1103/PhysRevB.99.174509} {\bibfield  {journal} {\bibinfo  {journal} {Phys.
  Rev. B}\ }\textbf {\bibinfo {volume} {99}},\ \bibinfo {pages} {174509}
  (\bibinfo {year} {2019})}\BibitemShut {NoStop}%
\bibitem [{\citenamefont {Saira}\ \emph {et~al.}(2012)\citenamefont {Saira},
  \citenamefont {Kemppinen}, \citenamefont {Maisi},\ and\ \citenamefont
  {Pekola}}]{Saira2012}%
  \BibitemOpen
  \bibfield  {author} {\bibinfo {author} {\bibfnamefont {O.~P.}\ \bibnamefont
  {Saira}}, \bibinfo {author} {\bibfnamefont {A.}~\bibnamefont {Kemppinen}},
  \bibinfo {author} {\bibfnamefont {V.~F.}\ \bibnamefont {Maisi}}, \ and\
  \bibinfo {author} {\bibfnamefont {J.~P.}\ \bibnamefont {Pekola}},\ }\bibfield
   {title} {\enquote {\bibinfo {title} {{Vanishing quasiparticle density in a
  hybrid Al/Cu/Al single-electron transistor}},}\ }\href {\doibase
  10.1103/PhysRevB.85.012504} {\bibfield  {journal} {\bibinfo  {journal} {Phys.
  Rev. B}\ }\textbf {\bibinfo {volume} {85}},\ \bibinfo {pages} {1} (\bibinfo
  {year} {2012})}\BibitemShut {NoStop}%
\bibitem [{\citenamefont {Chang}\ and\ \citenamefont
  {Scalapino}(1978)}]{Chang1978}%
  \BibitemOpen
  \bibfield  {author} {\bibinfo {author} {\bibfnamefont {J.~J.}\ \bibnamefont
  {Chang}}\ and\ \bibinfo {author} {\bibfnamefont {D.~J.}\ \bibnamefont
  {Scalapino}},\ }\bibfield  {title} {\enquote {\bibinfo {title}
  {{Nonequilibrium superconductivity}},}\ }\href {\doibase 10.1007/BF00116228}
  {\bibfield  {journal} {\bibinfo  {journal} {J. Low Temp. Phys.}\ }\textbf
  {\bibinfo {volume} {31}},\ \bibinfo {pages} {1} (\bibinfo {year}
  {1978})}\BibitemShut {NoStop}%
\bibitem [{\citenamefont {Sentef}\ \emph {et~al.}(2016)\citenamefont {Sentef},
  \citenamefont {Kemper}, \citenamefont {Georges},\ and\ \citenamefont
  {Kollath}}]{Sentef2016}%
  \BibitemOpen
  \bibfield  {author} {\bibinfo {author} {\bibfnamefont {M.~A.}\ \bibnamefont
  {Sentef}}, \bibinfo {author} {\bibfnamefont {A.~F.}\ \bibnamefont {Kemper}},
  \bibinfo {author} {\bibfnamefont {A.}~\bibnamefont {Georges}}, \ and\
  \bibinfo {author} {\bibfnamefont {C.}~\bibnamefont {Kollath}},\ }\bibfield
  {title} {\enquote {\bibinfo {title} {Theory of light-enhanced phonon-mediated
  superconductivity},}\ }\href {\doibase 10.1103/PhysRevB.93.144506} {\bibfield
   {journal} {\bibinfo  {journal} {Phys. Rev. B}\ }\textbf {\bibinfo {volume}
  {93}},\ \bibinfo {pages} {144506} (\bibinfo {year} {2016})}\BibitemShut
  {NoStop}%
\bibitem [{\citenamefont {Jauho}\ \emph {et~al.}(1994)\citenamefont {Jauho},
  \citenamefont {Wingreen},\ and\ \citenamefont {Meir}}]{Antipekka1994}%
  \BibitemOpen
  \bibfield  {author} {\bibinfo {author} {\bibfnamefont {A.-P.}\ \bibnamefont
  {Jauho}}, \bibinfo {author} {\bibfnamefont {N.~S.}\ \bibnamefont {Wingreen}},
  \ and\ \bibinfo {author} {\bibfnamefont {Y.}~\bibnamefont {Meir}},\
  }\bibfield  {title} {\enquote {\bibinfo {title} {Time-dependent transport in
  interacting and noninteracting resonant-tunneling systems},}\ }\href
  {\doibase 10.1103/PhysRevB.50.5528} {\bibfield  {journal} {\bibinfo
  {journal} {Phys. Rev. B}\ }\textbf {\bibinfo {volume} {50}},\ \bibinfo
  {pages} {5528} (\bibinfo {year} {1994})}\BibitemShut {NoStop}%
\bibitem [{\citenamefont {Pastawski}(1992)}]{Horacio1992}%
  \BibitemOpen
  \bibfield  {author} {\bibinfo {author} {\bibfnamefont {H.~M.}\ \bibnamefont
  {Pastawski}},\ }\bibfield  {title} {\enquote {\bibinfo {title} {Classical and
  quantum transport from generalized landauer-b\"uttiker equations. ii.
  time-dependent resonant tunneling},}\ }\href {\doibase
  10.1103/PhysRevB.46.4053} {\bibfield  {journal} {\bibinfo  {journal} {Phys.
  Rev. B}\ }\textbf {\bibinfo {volume} {46}},\ \bibinfo {pages} {4053}
  (\bibinfo {year} {1992})}\BibitemShut {NoStop}%
\bibitem [{\citenamefont {Xu}\ \emph {et~al.}(2019)\citenamefont {Xu},
  \citenamefont {Morimoto}, \citenamefont {Lanzara},\ and\ \citenamefont
  {Moore}}]{Moore2019}%
  \BibitemOpen
  \bibfield  {author} {\bibinfo {author} {\bibfnamefont {T.}~\bibnamefont
  {Xu}}, \bibinfo {author} {\bibfnamefont {T.}~\bibnamefont {Morimoto}},
  \bibinfo {author} {\bibfnamefont {A.}~\bibnamefont {Lanzara}}, \ and\
  \bibinfo {author} {\bibfnamefont {J.~E.}\ \bibnamefont {Moore}},\ }\bibfield
  {title} {\enquote {\bibinfo {title} {Efficient prediction of time- and
  angle-resolved photoemission spectroscopy measurements on a nonequilibrium
  bcs superconductor},}\ }\href {\doibase 10.1103/PhysRevB.99.035117}
  {\bibfield  {journal} {\bibinfo  {journal} {Phys. Rev. B}\ }\textbf {\bibinfo
  {volume} {99}},\ \bibinfo {pages} {035117} (\bibinfo {year}
  {2019})}\BibitemShut {NoStop}%
\bibitem [{\citenamefont {Kennes}\ \emph {et~al.}(2017)\citenamefont {Kennes},
  \citenamefont {Wilner}, \citenamefont {Reichman},\ and\ \citenamefont
  {Millis}}]{Millis2017}%
  \BibitemOpen
  \bibfield  {author} {\bibinfo {author} {\bibfnamefont {D.~M.}\ \bibnamefont
  {Kennes}}, \bibinfo {author} {\bibfnamefont {E.~Y.}\ \bibnamefont {Wilner}},
  \bibinfo {author} {\bibfnamefont {D.~R.}\ \bibnamefont {Reichman}}, \ and\
  \bibinfo {author} {\bibfnamefont {A.~J.}\ \bibnamefont {Millis}},\ }\bibfield
   {title} {\enquote {\bibinfo {title} {Nonequilibrium optical conductivity:
  General theory and application to transient phases},}\ }\href {\doibase
  10.1103/PhysRevB.96.054506} {\bibfield  {journal} {\bibinfo  {journal} {Phys.
  Rev. B}\ }\textbf {\bibinfo {volume} {96}},\ \bibinfo {pages} {054506}
  (\bibinfo {year} {2017})}\BibitemShut {NoStop}%
\bibitem [{\citenamefont {Cea}\ \emph {et~al.}(2018)\citenamefont {Cea},
  \citenamefont {Barone}, \citenamefont {Castellani},\ and\ \citenamefont
  {Benfatto}}]{Cea2018}%
  \BibitemOpen
  \bibfield  {author} {\bibinfo {author} {\bibfnamefont {T.}~\bibnamefont
  {Cea}}, \bibinfo {author} {\bibfnamefont {P.}~\bibnamefont {Barone}},
  \bibinfo {author} {\bibfnamefont {C.}~\bibnamefont {Castellani}}, \ and\
  \bibinfo {author} {\bibfnamefont {L.}~\bibnamefont {Benfatto}},\ }\bibfield
  {title} {\enquote {\bibinfo {title} {{Polarization dependence of the
  third-harmonic generation in multiband superconductors}},}\ }\href {\doibase
  10.1103/PhysRevB.97.094516} {\bibfield  {journal} {\bibinfo  {journal} {Phys.
  Rev. B}\ }\textbf {\bibinfo {volume} {97}},\ \bibinfo {pages} {094516}
  (\bibinfo {year} {2018})}\BibitemShut {NoStop}%
\bibitem [{\citenamefont {Ferr\'on}\ \emph {et~al.}(2012)\citenamefont
  {Ferr\'on}, \citenamefont {Dom\'{\i}nguez},\ and\ \citenamefont
  {S\'anchez}}]{Ferron2012}%
  \BibitemOpen
  \bibfield  {author} {\bibinfo {author} {\bibfnamefont {A.}~\bibnamefont
  {Ferr\'on}}, \bibinfo {author} {\bibfnamefont {D.}~\bibnamefont
  {Dom\'{\i}nguez}}, \ and\ \bibinfo {author} {\bibfnamefont {M.~J.}\
  \bibnamefont {S\'anchez}},\ }\bibfield  {title} {\enquote {\bibinfo {title}
  {Tailoring population inversion in landau-zener-st\"uckelberg interferometry
  of flux qubits},}\ }\href {\doibase 10.1103/PhysRevLett.109.237005}
  {\bibfield  {journal} {\bibinfo  {journal} {Phys. Rev. Lett.}\ }\textbf
  {\bibinfo {volume} {109}},\ \bibinfo {pages} {237005} (\bibinfo {year}
  {2012})}\BibitemShut {NoStop}%
\bibitem [{\citenamefont {Freericks}\ \emph {et~al.}(2009)\citenamefont
  {Freericks}, \citenamefont {Krishnamurthy},\ and\ \citenamefont
  {Pruschke}}]{Freericks2009}%
  \BibitemOpen
  \bibfield  {author} {\bibinfo {author} {\bibfnamefont {J.~K.}\ \bibnamefont
  {Freericks}}, \bibinfo {author} {\bibfnamefont {H.~R.}\ \bibnamefont
  {Krishnamurthy}}, \ and\ \bibinfo {author} {\bibfnamefont {T.}~\bibnamefont
  {Pruschke}},\ }\bibfield  {title} {\enquote {\bibinfo {title} {Theoretical
  description of time-resolved photoemission spectroscopy: Application to
  pump-probe experiments},}\ }\href {\doibase 10.1103/PhysRevLett.102.136401}
  {\bibfield  {journal} {\bibinfo  {journal} {Phys. Rev. Lett.}\ }\textbf
  {\bibinfo {volume} {102}},\ \bibinfo {pages} {136401} (\bibinfo {year}
  {2009})}\BibitemShut {NoStop}%
\bibitem [{\citenamefont {Shirley}(1965)}]{Shirley65}%
  \BibitemOpen
  \bibfield  {author} {\bibinfo {author} {\bibfnamefont {J.~H.}\ \bibnamefont
  {Shirley}},\ }\bibfield  {title} {\enquote {\bibinfo {title} {Solution of the
  schr\"odinger equation with a hamiltonian periodic in time},}\ }\href
  {\doibase 10.1103/PhysRev.138.B979} {\bibfield  {journal} {\bibinfo
  {journal} {Phys. Rev.}\ }\textbf {\bibinfo {volume} {138}},\ \bibinfo {pages}
  {B979} (\bibinfo {year} {1965})}\BibitemShut {NoStop}%
\bibitem [{\citenamefont {Sambe}(1973)}]{Sambe73}%
  \BibitemOpen
  \bibfield  {author} {\bibinfo {author} {\bibfnamefont {H.}~\bibnamefont
  {Sambe}},\ }\bibfield  {title} {\enquote {\bibinfo {title} {Steady states and
  quasienergies of a quantum-mechanical system in an oscillating field},}\
  }\href {\doibase 10.1103/PhysRevA.7.2203} {\bibfield  {journal} {\bibinfo
  {journal} {Phys. Rev. A}\ }\textbf {\bibinfo {volume} {7}},\ \bibinfo {pages}
  {2203} (\bibinfo {year} {1973})}\BibitemShut {NoStop}%
\bibitem [{\citenamefont {Kohler}\ \emph {et~al.}(2005)\citenamefont {Kohler},
  \citenamefont {Lehmann},\ and\ \citenamefont {Hänggi}}]{Kohler2005}%
  \BibitemOpen
  \bibfield  {author} {\bibinfo {author} {\bibfnamefont {S.}~\bibnamefont
  {Kohler}}, \bibinfo {author} {\bibfnamefont {J.}~\bibnamefont {Lehmann}}, \
  and\ \bibinfo {author} {\bibfnamefont {P.}~\bibnamefont {Hänggi}},\
  }\bibfield  {title} {\enquote {\bibinfo {title} {Driven quantum transport on
  the nanoscale},}\ }\href {\doibase
  https://doi.org/10.1016/j.physrep.2004.11.002} {\bibfield  {journal}
  {\bibinfo  {journal} {Physics Reports}\ }\textbf {\bibinfo {volume} {406}},\
  \bibinfo {pages} {379 } (\bibinfo {year} {2005})}\BibitemShut {NoStop}%
\bibitem [{\citenamefont {Grifoni}\ and\ \citenamefont
  {Hanggi}(1998)}]{Grifoni98}%
  \BibitemOpen
  \bibfield  {author} {\bibinfo {author} {\bibfnamefont {M.}~\bibnamefont
  {Grifoni}}\ and\ \bibinfo {author} {\bibfnamefont {P.}~\bibnamefont
  {Hanggi}},\ }\bibfield  {title} {\enquote {\bibinfo {title} {Driven quantum
  tunneling},}\ }\href {\doibase https://doi.org/10.1016/S0370-1573(98)00022-2}
  {\bibfield  {journal} {\bibinfo  {journal} {Physics Reports}\ }\textbf
  {\bibinfo {volume} {304}},\ \bibinfo {pages} {229 } (\bibinfo {year}
  {1998})}\BibitemShut {NoStop}%
\bibitem [{\citenamefont {Peralta~Gavensky}\ \emph {et~al.}(2018)\citenamefont
  {Peralta~Gavensky}, \citenamefont {Usaj},\ and\ \citenamefont
  {Balseiro}}]{Lucila2018}%
  \BibitemOpen
  \bibfield  {author} {\bibinfo {author} {\bibfnamefont {L.}~\bibnamefont
  {Peralta~Gavensky}}, \bibinfo {author} {\bibfnamefont {G.}~\bibnamefont
  {Usaj}}, \ and\ \bibinfo {author} {\bibfnamefont {C.~A.}\ \bibnamefont
  {Balseiro}},\ }\bibfield  {title} {\enquote {\bibinfo {title} {Time-resolved
  hall conductivity of pulse-driven topological quantum systems},}\ }\href
  {\doibase 10.1103/PhysRevB.98.165414} {\bibfield  {journal} {\bibinfo
  {journal} {Phys. Rev. B}\ }\textbf {\bibinfo {volume} {98}},\ \bibinfo
  {pages} {165414} (\bibinfo {year} {2018})}\BibitemShut {NoStop}%
\bibitem [{\citenamefont {Morgenstern}(2010)}]{Morgenstern1609}%
  \BibitemOpen
  \bibfield  {author} {\bibinfo {author} {\bibfnamefont {M.}~\bibnamefont
  {Morgenstern}},\ }\bibfield  {title} {\enquote {\bibinfo {title} {{STM} ready
  for the time domain},}\ }\href {\doibase 10.1126/science.1194918} {\bibfield
  {journal} {\bibinfo  {journal} {Science}\ }\textbf {\bibinfo {volume}
  {329}},\ \bibinfo {pages} {1609} (\bibinfo {year} {2010})}\BibitemShut
  {NoStop}%
\bibitem [{\citenamefont {Nunes}\ and\ \citenamefont
  {Freeman}(1993)}]{Nunes1029}%
  \BibitemOpen
  \bibfield  {author} {\bibinfo {author} {\bibfnamefont {G.}~\bibnamefont
  {Nunes}}\ and\ \bibinfo {author} {\bibfnamefont {M.~R.}\ \bibnamefont
  {Freeman}},\ }\bibfield  {title} {\enquote {\bibinfo {title} {Picosecond
  resolution in scanning tunneling microscopy},}\ }\href {\doibase
  10.1126/science.262.5136.1029} {\bibfield  {journal} {\bibinfo  {journal}
  {Science}\ }\textbf {\bibinfo {volume} {262}},\ \bibinfo {pages} {1029}
  (\bibinfo {year} {1993})}\BibitemShut {NoStop}%
\bibitem [{\citenamefont {Takeuchi}\ \emph {et~al.}(2004)\citenamefont
  {Takeuchi}, \citenamefont {Aoyama}, \citenamefont {Oshima}, \citenamefont
  {Okada}, \citenamefont {Oigawa}, \citenamefont {Sano}, \citenamefont
  {Shigekawa}, \citenamefont {Morita},\ and\ \citenamefont
  {Yamashita}}]{Mikio2004}%
  \BibitemOpen
  \bibfield  {author} {\bibinfo {author} {\bibfnamefont {O.}~\bibnamefont
  {Takeuchi}}, \bibinfo {author} {\bibfnamefont {M.}~\bibnamefont {Aoyama}},
  \bibinfo {author} {\bibfnamefont {R.}~\bibnamefont {Oshima}}, \bibinfo
  {author} {\bibfnamefont {Y.}~\bibnamefont {Okada}}, \bibinfo {author}
  {\bibfnamefont {H.}~\bibnamefont {Oigawa}}, \bibinfo {author} {\bibfnamefont
  {N.}~\bibnamefont {Sano}}, \bibinfo {author} {\bibfnamefont {H.}~\bibnamefont
  {Shigekawa}}, \bibinfo {author} {\bibfnamefont {R.}~\bibnamefont {Morita}}, \
  and\ \bibinfo {author} {\bibfnamefont {M.}~\bibnamefont {Yamashita}},\
  }\bibfield  {title} {\enquote {\bibinfo {title} {Probing subpicosecond
  dynamics using pulsed laser combined scanning tunneling microscopy},}\ }\href
  {\doibase 10.1063/1.1804238} {\bibfield  {journal} {\bibinfo  {journal}
  {Applied Physics Letters}\ }\textbf {\bibinfo {volume} {85}},\ \bibinfo
  {pages} {3268} (\bibinfo {year} {2004})}\BibitemShut {NoStop}%
\bibitem [{\citenamefont {{Zmuidzinas}}\ and\ \citenamefont
  {{LeDuc}}(1992)}]{Zmuidzinas1992}%
  \BibitemOpen
  \bibfield  {author} {\bibinfo {author} {\bibfnamefont {J.}~\bibnamefont
  {{Zmuidzinas}}}\ and\ \bibinfo {author} {\bibfnamefont {H.~G.}\ \bibnamefont
  {{LeDuc}}},\ }\bibfield  {title} {\enquote {\bibinfo {title} {Quasi-optical
  slot antenna sis mixers},}\ }\href {\doibase 10.1109/22.156607} {\bibfield
  {journal} {\bibinfo  {journal} {IEEE Transactions on Microwave Theory and
  Techniques}\ }\textbf {\bibinfo {volume} {40}},\ \bibinfo {pages} {1797}
  (\bibinfo {year} {1992})}\BibitemShut {NoStop}%
\bibitem [{\citenamefont {Leridon}(1997)}]{Leridon1997}%
  \BibitemOpen
  \bibfield  {author} {\bibinfo {author} {\bibfnamefont {B.}~\bibnamefont
  {Leridon}},\ }\bibfield  {title} {\enquote {\bibinfo {title} {Josephson noise
  in quasiparticle mixers},}\ }\href {\doibase 10.1063/1.364354} {\bibfield
  {journal} {\bibinfo  {journal} {Journal of Applied Physics}\ }\textbf
  {\bibinfo {volume} {81}},\ \bibinfo {pages} {3243} (\bibinfo {year}
  {1997})}\BibitemShut {NoStop}%
\bibitem [{\citenamefont {Mikhailovsky}\ \emph {et~al.}(1991)\citenamefont
  {Mikhailovsky}, \citenamefont {Shulga}, \citenamefont {Karakozov},
  \citenamefont {Dolgov},\ and\ \citenamefont {Maksimov}}]{Mikhailovsky1991}%
  \BibitemOpen
  \bibfield  {author} {\bibinfo {author} {\bibfnamefont {A.}~\bibnamefont
  {Mikhailovsky}}, \bibinfo {author} {\bibfnamefont {S.}~\bibnamefont
  {Shulga}}, \bibinfo {author} {\bibfnamefont {A.}~\bibnamefont {Karakozov}},
  \bibinfo {author} {\bibfnamefont {O.}~\bibnamefont {Dolgov}}, \ and\ \bibinfo
  {author} {\bibfnamefont {E.}~\bibnamefont {Maksimov}},\ }\bibfield  {title}
  {\enquote {\bibinfo {title} {Thermal pair-breaking in superconductors with
  strong electron-phonon interaction},}\ }\href {\doibase
  https://doi.org/10.1016/0038-1098(91)90062-Z} {\bibfield  {journal} {\bibinfo
   {journal} {Solid State Communications}\ }\textbf {\bibinfo {volume} {80}},\
  \bibinfo {pages} {511 } (\bibinfo {year} {1991})}\BibitemShut {NoStop}%
\bibitem [{\citenamefont {Pekola}\ \emph {et~al.}(2010)\citenamefont {Pekola},
  \citenamefont {Maisi}, \citenamefont {Kafanov}, \citenamefont {Chekurov},
  \citenamefont {Kemppinen}, \citenamefont {Pashkin}, \citenamefont {Saira},
  \citenamefont {M\"ott\"onen},\ and\ \citenamefont {Tsai}}]{Pekola2010}%
  \BibitemOpen
  \bibfield  {author} {\bibinfo {author} {\bibfnamefont {J.~P.}\ \bibnamefont
  {Pekola}}, \bibinfo {author} {\bibfnamefont {V.~F.}\ \bibnamefont {Maisi}},
  \bibinfo {author} {\bibfnamefont {S.}~\bibnamefont {Kafanov}}, \bibinfo
  {author} {\bibfnamefont {N.}~\bibnamefont {Chekurov}}, \bibinfo {author}
  {\bibfnamefont {A.}~\bibnamefont {Kemppinen}}, \bibinfo {author}
  {\bibfnamefont {Y.~A.}\ \bibnamefont {Pashkin}}, \bibinfo {author}
  {\bibfnamefont {O.-P.}\ \bibnamefont {Saira}}, \bibinfo {author}
  {\bibfnamefont {M.}~\bibnamefont {M\"ott\"onen}}, \ and\ \bibinfo {author}
  {\bibfnamefont {J.~S.}\ \bibnamefont {Tsai}},\ }\bibfield  {title} {\enquote
  {\bibinfo {title} {Environment-assisted tunneling as an origin of the dynes
  density of states},}\ }\href {\doibase 10.1103/PhysRevLett.105.026803}
  {\bibfield  {journal} {\bibinfo  {journal} {Phys. Rev. Lett.}\ }\textbf
  {\bibinfo {volume} {105}},\ \bibinfo {pages} {026803} (\bibinfo {year}
  {2010})}\BibitemShut {NoStop}%
\bibitem [{\citenamefont {Herman}\ and\ \citenamefont
  {Hlubina}(2016)}]{Hlubina2016}%
  \BibitemOpen
  \bibfield  {author} {\bibinfo {author} {\bibfnamefont {F.~c.~v.}\
  \bibnamefont {Herman}}\ and\ \bibinfo {author} {\bibfnamefont
  {R.}~\bibnamefont {Hlubina}},\ }\bibfield  {title} {\enquote {\bibinfo
  {title} {Microscopic interpretation of the dynes formula for the tunneling
  density of states},}\ }\href {\doibase 10.1103/PhysRevB.94.144508} {\bibfield
   {journal} {\bibinfo  {journal} {Phys. Rev. B}\ }\textbf {\bibinfo {volume}
  {94}},\ \bibinfo {pages} {144508} (\bibinfo {year} {2016})}\BibitemShut
  {NoStop}%
\bibitem [{\citenamefont {Herman}\ and\ \citenamefont
  {Hlubina}(2018)}]{Hlubina2018}%
  \BibitemOpen
  \bibfield  {author} {\bibinfo {author} {\bibfnamefont {F.~c.~v.}\
  \bibnamefont {Herman}}\ and\ \bibinfo {author} {\bibfnamefont
  {R.}~\bibnamefont {Hlubina}},\ }\bibfield  {title} {\enquote {\bibinfo
  {title} {Thermodynamic properties of dynes superconductors},}\ }\href
  {\doibase 10.1103/PhysRevB.97.014517} {\bibfield  {journal} {\bibinfo
  {journal} {Phys. Rev. B}\ }\textbf {\bibinfo {volume} {97}},\ \bibinfo
  {pages} {014517} (\bibinfo {year} {2018})}\BibitemShut {NoStop}%
\end{thebibliography}%

\end{document}